\documentclass[final,5p,twocolumn,sort&compress,super]{elsarticle}

\usepackage[justification=centering,font=small]{caption}
\usepackage{float}
\usepackage{geometry}
\usepackage{natbib}
\usepackage{setspace}
\usepackage{xkeyval}

\usepackage{fancyhdr}
\makeatletter
\def\ps@pprintTitle{%
  \let\@oddhead\@empty
  \let\@evenhead\@empty
  \def\@oddfoot{\reset@font\slshape{Accepted for publication in The Journal of Physical Chemistry A}\hfil\slshape{26 July 2016}}
  \let\@evenfoot\@oddfoot
}
\makeatother

\usepackage[version=3]{mhchem} 

\usepackage[T1]{fontenc}       
\usepackage{tgschola}

\usepackage{titlesec}

\titleformat*{\section}{\large\bfseries}
\titleformat*{\subsection}{\normalsize\bfseries}

\usepackage{graphicx}
\usepackage[figuresright]{rotating}
\usepackage{booktabs}
\usepackage{footnote}
\usepackage[hang,splitrule]{footmisc}
\usepackage{amsmath,amsfonts,mathabx}
\usepackage{balance}

\makeatletter
\newcommand\footnoteref[1]{\protected@xdef\@thefnmark{\ref{#1}}\@footnotemark}
\makeatother

\makeatletter
\DeclareRobustCommand*{\bfseries}{%
  \not@math@alphabet\bfseries\mathbf
  \fontseries\bfdefault\selectfont
  \boldmath
}
\makeatother

\newcommand{\wn}{\ensuremath{\mathrm{cm}^{-1}}}

\usepackage[driverfallback=dvipdfm]{hyperref}
\usepackage{xcolor}
\hypersetup{
    colorlinks,
    linkcolor={red!50!black},
    citecolor={blue!50!black},
    urlcolor={blue!80!black}
}

\begin{document}

\begin{frontmatter}

\title{A Spectroscopic Survey of Electronic Transitions of \ce{C6H}, \ce{^{13}C6H}, and \ce{C6D}}


\author[1,2]{X. Bacalla}

\author[1,3]{E.J. Salumbides}

\author[2]{H. Linnartz}

\author[1]{W. Ubachs}

\author[1,2]{D. Zhao\fnref{presentadd}\corref{correspondence}}
\cortext[correspondence]{Corresponding author.}
\ead{dzhao@ustc.edu.cn}
\fntext[presentadd]{Present address: Hefei National Laboratory for Physical Sciences at the Microscale, Department of Chemical Physics, University of Science and Technology of China, 96~Jinzhai~Road, Hefei, Anhui 230026, P. R. China.}

\address[1]{Department of Physics and Astronomy, LaserLaB, VU University Amsterdam, De~Boelelaan 1081, NL-1081 HV Amsterdam, The Netherlands}

\address[2]{Sackler Laboratory for Astrophysics, Leiden Observatory, Leiden University, PO~Box~9513, NL-2300 RA Leiden, The Netherlands}

\address[3]{Department of Physics, University of San Carlos, Nasipit, Talamban, Cebu City 6000, Philippines}


\renewcommand{\abstractname}{\Large ABSTRACT}
\begin{abstract}
Electronic spectra of \ce{C6H} are measured in the $18\,950 - 21\,100$~\wn\ domain using cavity ring-down spectroscopy of a supersonically expanding hydrocarbon plasma. In total, 19 (sub)bands of \ce{C6H} are presented, all probing the vibrational manifold of the B$^2\Pi$ electronically excited state. The assignments are guided by electronic spectra available from matrix isolation work, isotopic substitution experiments (yielding also spectra for \ce{^{13}C6H} and \ce{C6D}), predictions from \emph{ab initio} calculations as well as rotational fitting and vibrational contour simulations using the available ground state parameters as obtained from microwave experiments. 
Besides the $0^0_0$ origin band, three non-degenerate stretching vibrations along the linear backbone of the \ce{C6H} molecule are assigned: the $\nu_6$ mode associated with the C--C bond vibration and the $\nu_4$ and $\nu_3$ modes associated with C$\equiv$C triple bonds. For the two lowest $\nu_{11}$ and $\nu_{10}$ bending modes, a Renner-Teller analysis is performed identifying the $\mu^2\Sigma$($\nu_{11}$) and both $\mu^2\Sigma$($\nu_{10}$) and $\kappa^2\Sigma$($\nu_{10}$) components. In addition, two higher lying bending modes are observed, which are tentatively assigned as  $\mu^2\Sigma$($\nu_9$)  and $\mu^2\Sigma$($\nu_8$) levels. In the excitation region below the first non-degenerate vibration ($\nu_6$), some $^2\Pi - {^2}\Pi$ transitions are observed that are assigned as even combination modes of low-lying bending vibrations. The same holds for a $^2\Pi - {^2}\Pi$ transition found above the $\nu_6$ level. From these spectroscopic data and the vibronic analysis a comprehensive energy level diagram for the B$^2\Pi$ state of \ce{C6H} is derived and presented.
\end{abstract}

\begin{keyword}
hexatriynyl radical \sep cavity ring-down spectroscopy \sep isotopic substitution \sep supersonic plasma discharge


\end{keyword}

\end{frontmatter}


\section*{Introduction}
\label{Introduction}
\vspace{-3pt}
The hexatriynyl radical, \ce{C6H}, belongs to the best studied linear carbon chain radicals. As several other members in the acetylenic C$_{2n}$H series, it has been identified in different environments in the interstellar medium.~\cite{Suzuki1986,Cernicharo1987b,Guelin1987,Kawaguchi1995} Microwave spectroscopic work~\cite{Pearson1988} has provided accurate spectroscopic constants for the X$^2{\Pi}$ electronic ground state that later were extended to deuterated hexatriynyl~\cite{Linnartz1999} (\ce{C6D}) as well as to vibrationally excited \ce{C6H}.~\cite{Gottlieb2010} \emph{Ab initio} calculations,~\cite{Liu1992,Brown1999,Cao2001} together with infrared (IR) work,~\cite{Doyle1991} have also been conducted to predict and identify the fundamental vibrational frequencies of \ce{C6H}. Recently, the astronomical detection of vibrationally excited \ce{C6H}~\cite{Cernicharo2008,Gottlieb2010} and the chemically related \ce{C6H-} anion~\cite{McCarthy2006} triggered a renewed spectroscopic interest in this molecule.

The first electronic spectra of \ce{C6H} were recorded using matrix isolation spectroscopy;~\cite{Forney1995,Freivogel1995} results of which guided two gas phase studies.~\cite{Kotterer1997,Linnartz1999} An unresolved electronic origin band as well as several hot bands were recorded for a rather high temperature ($T_\mathrm{rot}\sim$~150~K) using a liquid-nitrogen cooled hollow cathode discharge.~\cite{Kotterer1997} After this, a study of a supersonically expanding and adiabatically cooled plasma resulted in a full rotational analysis of the two spin-orbit components of the $^2{\Pi}-$X$^2{\Pi}$ origin band spectrum for both \ce{C6H} and \ce{C6D} radicals.~\cite{Linnartz1999} An extended study, presenting more accurate data for the electronic origin band as well as a Renner-Teller analysis for the $\{11\}_1^1$ ${\mu}^2{\Sigma}-{\mu}^2{\Sigma}$ vibronic hot band, was presented several years later.~\cite{Zhao2011b} In the optical spectra only the $^2\Sigma$ vibronic component of the $\nu_{11}$ mode was observed, in contrast to the $^2\Delta$ component, which was detected in the microwave spectrum,~\cite{Gottlieb2010} but which is most likely too energetic ($\sim$~100~\wn) to be observed in cold jet expansions. The same holds for the low-lying electronic state with $^2\Sigma$ symmetry, at an excitation energy of $\sim$~$1\,774$~\wn, which was observed in a photo-detachment study.~\cite{Taylor1998} These states were predicted in \emph{ab initio} electronic structure calculations.~\cite{Sobolewski1995,Cao2001} Harmonic vibrational frequencies of the various modes in \ce{C6H} were \emph{ab initio} calculated, yielding the lowest mode, $\nu_{11}$, at some $100$~\wn\ and the highest mode at about $3\,300$~\wn.~\cite{Brown1999} Also of interest is the spectroscopic investigation of the \ce{C6H+} cation, both in a neon matrix~\cite{Shnitko2006} and in the gas phase.~\cite{Raghunandan2010}

In the present work, cavity ring-down spectroscopy of a supersonically expanding hydrocarbon plasma has been employed in a wider-ranging spectroscopic survey. All spectroscopic features observed are associated with the B$^2{\Pi}$ electronically excited state and its vibrational modes. From the \ce{C6H} origin band at $18\,980$~\wn\ towards higher wavenumbers, several new spectral features have been observed and found to originate from \ce{C6H} as well. Also, a substantial number of \ce{^{13}C6H} (full \ce{^{13}C} substitution) and \ce{C6D} bands has been observed for the first time.

\section*{Experiment}
The experimental setup and the measurement procedure are similar to those described before.~\cite{Motylewski1999,Zhao2011b} The carbon chain radicals are formed by discharging different gas mixtures in the throat of a pulsed plasma nozzle, either a slit nozzle~\cite{Motylewski1999} or pinhole design.~\cite{Zhao2011} For C$_6$H and C$_6$D, 0.5\% $^{12}$C$_2$H$_2$ in He and 0.3\% $^{12}$C$_2$D$_2$ in He mixtures are used, respectively. An isotopically enriched $^{13}$C$_2$H$_2$ (99\% purity) gas sample, mixed at 0.13--0.18\% in He/Ar (85/15) is used to record fully \ce{^{13}C}-substituted $^{13}$C$_6$H spectra. A solenoid valve (General Valve, Series 9) on top of the nozzle body controls a high pressure (5--10~bar) gas pulse (roughly of 1~ms duration) that is discharged between ceramically isolated electrodes before expanding into a vacuum, generated by a powerful roots blower system. Typically, 300 to 500 $\mu$s long negative high voltage pulses (V/I $\sim$~$-$750~V / 100~mA) are offered to dissociate the precursor species and to allow molecule formation through collisions in the expanding plasma. In parallel the adiabatic expansion allows for final rotational temperatures in the order of 15 to 25~K.

The output of a pulsed Nd:YAG-pumped dye laser is aligned just below (2--5~mm) and parallel to a 3~cm $\times$ 300~$\mu$m slit nozzle, or across the conical jet of a 1.0~mm diameter pinhole. The effective absorption pathlength along the slit or through the pinhole expansion can be further improved by orders of magnitude by positioning the slit nozzle (or pinhole) along the optical axis between two highly reflective mirrors ($R$~>~0.99998) in a cavity ring-down configuration.~\cite{Wheeler1998} Typical values for the ring-down time amount to $\tau$ = 40 to 100~$\mu$s, corresponding to effective path lengths of several kilometers through the plasma. The experiment runs at 10~Hz and precise pulse generators are used to guarantee that gas and discharge pulse as well as ring-down event coincide in time. Typically the average of 10 ring-down events is used for each laser wavelength before being stored as a data point.

The laser bandwidth is about 0.035~cm$^{-1}$, thus sufficiently narrow to resolve subsequent rovibronic transitions in C$_6$H that has a rotational constant of the order of 0.045~cm$^{-1}$, i.e., two lines are separated by about 0.090~cm$^{-1}$. Wavelength calibration is achieved by recording simultaneously an etalon transmission spectrum (FSR $\sim$~20.1~GHz) which provides relative frequency markers for correction of any nonlinearity during scanning. An iodine (\ce{I2}) spectrum is recorded for absolute wavelength calibration, yielding an overall accuracy of better than 0.02~cm$^{-1}$.

The use of the slit nozzle leads to nearly Doppler-free spectra due to a collimation effect on the gas pulses emanating from the slit jet expansion perpendicular to the laser beam inside the ring-down cavity. In the case of the pinhole nozzle the contribution of molecules with velocity components along the laser beam path is larger, therewith producing spectral lines with extended Doppler broadening. However, the latter effect can be partially counteracted by mixing Ar in the gas sample, which produces a favorable effect on the spectral linewidth. This advantage is exploited for the recording of $^{13}$C$_6$H spectra. The isotopically enriched $^{13}$C$_2$H$_2$ was used in combination with the pinhole nozzle which uses up less gas than the slit nozzle. A few of the $^{13}$C-containing bands were nevertheless recorded in the slit nozzle configuration.

Obviously, the optical methods used are not mass selective, and spectra contain information from a number of species formed in the plasma, including separate transitions originating from, for example, C$_2$, C$_3$, or CH. Also, under similar experimental conditions, long linear \cite{Haddad2015} and bent chain \cite{Zhao2011c,Zhao2012} carbon-based molecules are formed in the plasma expansion. While under the presently used plasma expansion conditions the larger molecules are rotationally and vibrationally cooled, typically exhibiting rotational temperatures of $T_\mathrm{rot} \sim$~20~K and vibrational temperatures of the same values,~\cite{Zhao2011b} the diatomic radicals C$_2$ and CH are detected at elevated temperatures even allowing the detection of C$_2$ in the metastable a$^3\Pi_u$ state in the $v=9$ level.~\cite{Haddad2015} It is noted that the resulting rotational temperature can be regulated by changing the nozzle-to-laser-beam distance.~\cite{Zhao2011,Zhao2011b} Further, the plasma chemistry is different under conditions of a conical expansion when a pinhole nozzle is employed, or for a planar expansion in the case of a slit jet configuration. The spectral features associated with C$_2$ and CH, which in some cases can be assigned to previously identified transitions, overlap and in some cases congest the C$_6$H features under investigation in the present study. The isolated line features pertaining to C$_2$ and CH (and possibly other radicals) will be indicated by an asterisk ($*$). In particular, in the energy region higher than $19\,350$~\wn, with the (0,0) band origin of the d$^3\Pi_g -$a$^3\Pi_u$ \ce{C2} Swan band at $T_0 = 19\,378.4646(7)$~\wn,~\cite{Lloyd1999} the spectra become swamped by a line forest, for which reason no spectra for C$_6$D and $^{13}$C$_6$H will be reported in this range. For C$_6$H the sequence of vibronic states is presented despite the fact that the spectra are severely contaminated by C$_2$ lines. With the \texttt{PGOPHER}~\cite{PGopher} program and using spectroscopic constants from the literature,~\cite{Lloyd1999} we can simulate and identify most of the strong \ce{C2} lines originating from the (0,0) and (1,1) Swan bands.

In addition to the spectral features associated with the absorption lines of these diatomic radicals, some of the recorded spectra are overlaid by absorption features of unknown origin, which, due to their isolated nature, may be connected to atomic lines. These features, that will be indicated by hashes ($\#$), are in some cases detected as reduced absorptions, hinting at amplified spontaneous emission or laser-induced fluorescence features. Since we failed to assign those features they will be treated as artifacts in the following. We note here that the presently reported spectra were recorded over a 5-year interval (2010--2015) and a number of bands were fully reproduced using independently produced gas mixtures from which we conclude that these `artifacts' are not due to measurement issues but are reproducible spectral features associated with atomic or molecular absorption or emission, although unassigned.

\section*{Results}
The survey covers the $18\,950$--$21\,100$~\wn\ ($\sim$~527--473~nm) range and comprises a large number of bands. For each isotopologue, eight individual bands (seven for \ce{C6D}) have been unambiguously assigned.  In the case of \ce{C6H}, eleven additional bands have been recorded that exhibit very different spectral features. The assignments to C$_6$H, $^{13}$C$_6$H, and C$_6$D are based on a rotational analysis of the bands where it is assumed that in the cold jet expansion, the population is only retained in the lowest spin-doublet of the X$^2\Pi$ state for both $^2\Pi_{3/2}$ and $^2\Pi_{1/2}$ components and in the very low lying $\mu^2\Sigma$ Renner-Teller component of the $\nu_{11}$ bending vibration. For these ground states, accurate molecular constants are available from microwave data~\cite{Pearson1988,Linnartz1999,Gottlieb2010} except for the $^{13}$C$_6$H isotopologue for which the ground state constants were determined optically~\cite{Zhao2011b} and via isotopic scaling. The use of the available information allows us to identify the character of the lower state of the bands. As for the assignment of the excited state vibrational levels, this is based on neon matrix spectra \cite{Forney1995,Freivogel1995,Nagarajan2010} in combination with \emph{ab initio} calculations.~\cite{Doyle1991,Liu1992,Sobolewski1995,Brown1999,Cao2001}

\begin{figure}[]
  \centering
  \includegraphics[width=0.48\textwidth]{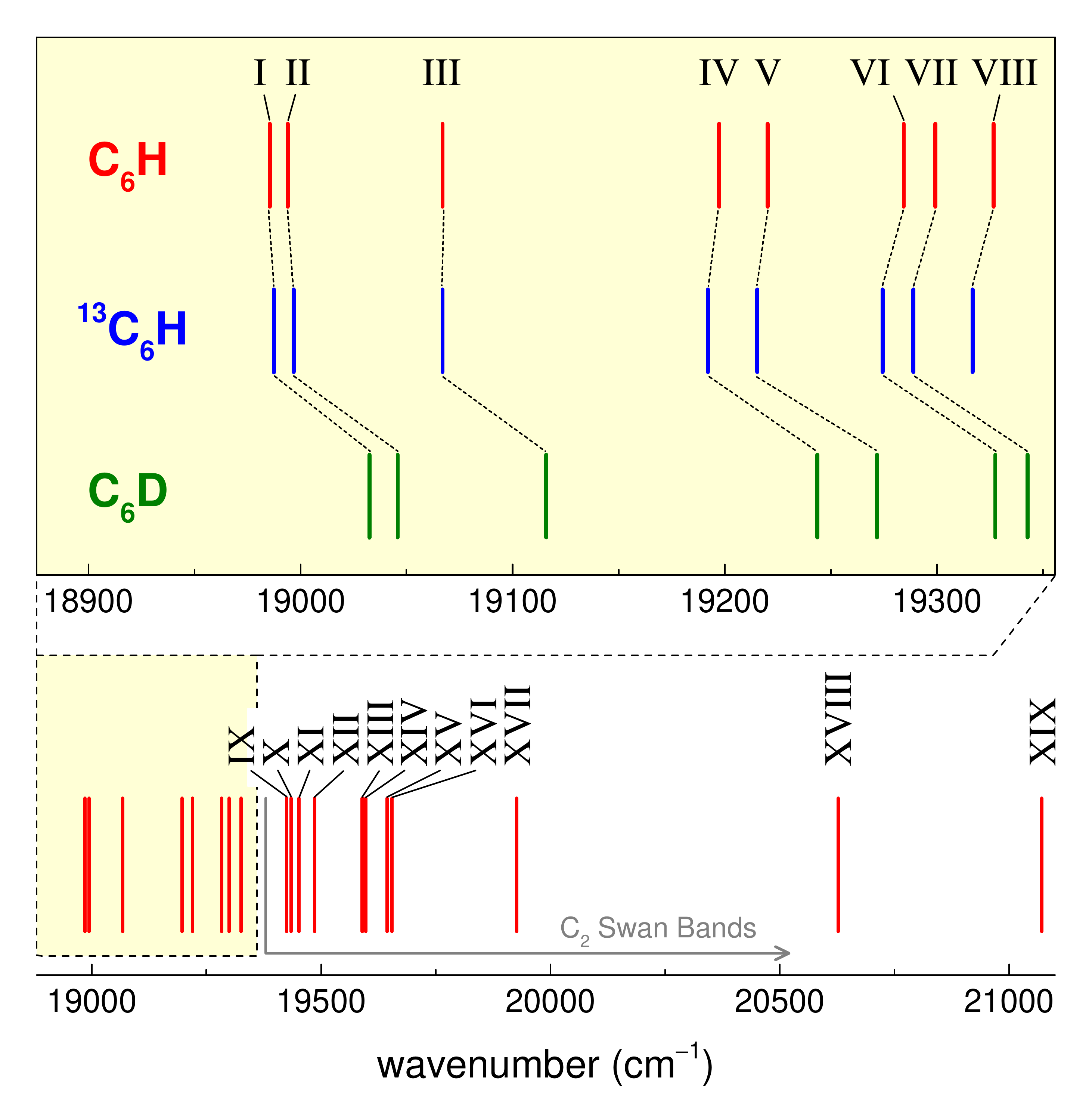}
  \caption{Stick diagram denoting the positions of all the observed bands in the $18\,950-21\,100$~\wn\ range for \ce{C6H} (in red, lower panel). The corresponding bands for \ce{^{13}C6H} (blue) and \ce{C6D} (green) are shown in the zoomed portion (upper panel) and are connected with dotted lines. The area with overlapping \ce{C2} Swan transitions is indicated.}
  \label{fig:compiled-allbands}
\end{figure}

In Fig.~\ref{fig:compiled-allbands} an overview of the band origins of all the observed optical absorption bands is shown for the three species studied here: regular C$_6$H, $^{13}$C$_6$H, and C$_6$D. Connected by dotted lines, the corresponding bands in the zoom-in (upper panel) of Fig.~\ref{fig:compiled-allbands} show a similar pattern for each of the three isotopologues. These isotopic shifts further guide in identifying the absorbing species. The labels with Roman numerals I--XIX in the figure correspond with those used in the sections below and in Table~\ref{tab:summary} and are based on the order at which the bands appear in the overall spectrum. A substantial number of the observed bands has been recorded at rotational resolution and for many of these bands accurate molecular parameters have been derived using \texttt{PGOPHER}. The bands were described either by a $^2\Pi-{^2}\Pi$ or a $^2\Sigma-{^2}\Sigma$ transition (see below). For the rotational and fine structure analysis, we adopt the effective Hamiltonian of a diatomic molecule, which may be used for transitions in a linear polyatomic molecule such as C$_6$H. The fitting routine takes ground state values for the rotational constant $B''$, the centrifugal distortion constant $D''$, the spin-orbit constant $A''$, and the spin-rotation constant $\gamma''$; these parameters are kept fixed in optimizing the excited state values. With the exception for the origin band (Bands I and II) and for the next vibronic band (Band III), we assume $D' \approx D''$: $1.35 \times 10^{-9}$~\wn\ (\ce{C6H}) and $1.2 \times 10^{-9}$~\wn\ (\ce{^{13}C6H} and \ce{C6D}). As in the plasma jet, particularly, lower lying rotational levels are probed, an independent fit of $D'$ is not possible for the majority of the bands presented here.

The resulting values are summarized in Table~\ref{tab:summary}. For some bands a rotational analysis could not be performed, either due to a poor signal-to-noise ratio, overlapping features from other species, or due to lifetime broadening. For these cases only approximate molecular constants are reported.

\begin{sidewaystable*}[h]
\centering
\caption{Summary of spectroscopic constants in cm$^{-1}$. Terminating the values (enclosed in parentheses) are the statistical uncertainties (1$\sigma$) obtained from a least-squares rotational line fit. For values obtained through a band contour simulation, the uncertainties are omitted. Numbers in italics are electronic ground state values taken from Refs. [\!\!\citenum{Linnartz1999}\,] and [\!\!\!\,\citenum{Gottlieb2010}\,] and are kept fixed in the fitting routine (with the exception for \ce{^{13}C6H}). Tentative mode assignments are enclosed in square brackets [\ ].}
\label{tab:summary}
\resizebox{\textwidth}{!}{%
\begin{tabular}{@{}clccccccccccc@{}}
\toprule
&     & $T_0$~\footnote{Resulting values from the fitting routine, but the absolute uncertainty is limited to 0.02~\wn\ from \ce{I2} calibration.}  & Band type                & $B'$            & $B''$              & $\gamma'$ ($10^{-3}$)  & $\gamma''$ ($10^{-3}$)  & $A'$  & $A''$      & Relative energy & Isotopic shift \footnote{Relative to the corresponding band in C$_6$H.} & Assignment \footnote{The notation, e.g. $\kappa\{10\}^1\ \mu\{11\}_1$, denotes a transition from the $\mu$ component of the $\nu_{11} = 1$ vibrational mode of the lower electronic state to the $\kappa$ component of the $\nu_{10} = 1$ mode of the upper electronic state. The mode numbering is adopted from Ref. [\!\!\citenum{Brown1999}\,].} \\
\midrule

C$_6$H&I,\,II\footnote{\label{foot:Zhao2011b}Data from Ref. [\!\!\citenum{Zhao2011b}\,].} & 18989.7672(4) & ${^2}\Pi-{^2}\Pi$ & 0.0455952(5) & \emph{0.04640497} & $-$3.63(18) & \emph{$-$7.12} & $-$23.69(2) & \emph{$-$15.04} & 0.0  & -      & $0_0^0$                 \\
&III\footnoteref{foot:Zhao2011b}  & 19066.8372(6)                                   & ${^2}\Sigma-{^2}\Sigma$  & 0.0456873(11)   & \emph{0.04652018}  & 38.29(6)    & \emph{$-$0.628}  & -               & -                  & 77.1    & -      & $\mu\{11\}_1^1$ bend           \\
&IV   & 19197.314(3)                                                                & ${^2}\Sigma-{^2}\Sigma$  & 0.045580(7)    & \emph{0.04652018}  & 79.55(18)   & \emph{$-$0.628}  & -               & -                  & 207.5   & -      & $\mu\{10\}^1\ \mu\{11\}_1$ bend \\
&V    & 19220.294(6)                                                                & ${^2}\Sigma-{^2}\Sigma$  & 0.045971(18)    & \emph{0.04652018}  & 107.1(4)    & \emph{$-$0.628}  & -               & -                  & 230.5   & -      & $\kappa\{10\}^1\ \mu\{11\}_1$ bend \\
&VI   & 19283.992(3)                                                                & ${^2}\Sigma-{^2}\Sigma$  & 0.045916(8)     & \emph{0.04652018}  & 70.27(16)   & \emph{$-$0.628}  & -               & -                  & 294.2   & -      & [\,$\mu \{9\}^1\ \mu\{11\}_1$\,] bend      \\
&VII  & 19298.56\footnote{\label{foot:contour}Value obtained from band contour simulation.}                                                                  & ${^2}\Pi-{^2}\Pi$        & 0.0457\footnoteref{foot:contour}      & \emph{0.04640497}  & 40\footnoteref{foot:contour}       & \emph{$-$7.12}   & $-$14.6\footnoteref{foot:contour}\footnote{\label{foot:tentative}Tentative value, derived from the assumption that the $^2\Pi_{1/2}$ component overlaps the $^2\Pi_{3/2}$ component.}     & \emph{$-$15.04}  & 308.8   & -      &                         \\
&VIII & 19326.479(4)                                                               & ${^2}\Pi-{^2}\Pi$        & 0.045392(11)     & \emph{0.04640497}  & -           & -                & -               & -                  & 336.7   & -      &       \\
&IX,\,X   & 19429.236(5)                                                            & ${^2}\Pi-{^2}\Pi$        & 0.046137(12)    & \emph{0.04640497}  & $-$7(6)       & \emph{$-$7.12}   & $-$23.77(2)    & \emph{$-$15.04}  & 439.5   & -      &                         \\
&XI   & 19451.792(2)                                                                & ${^2}\Pi-{^2}\Pi$        & 0.045969(5)     & \emph{0.04640497}  & -           & -                & -               & -                  & 462.0   & -      &                         \\
&XII  & 19486.08\footnoteref{foot:contour}                                                               & ${^2}\Sigma-{^2}\Sigma$  & 0.0455\footnoteref{foot:contour}      & \emph{0.04652018}  & 13\footnoteref{foot:contour}    & \emph{$-$0.628}  & -               & -                  & 496.3   & -      &     [\,$\mu \{8\}^1\ \mu\{11\}_1$\,] bend                    \\
&XIII,\,XIV & 19593.11\footnoteref{foot:contour}                                                            & ${^2}\Pi-{^2}\Pi$        & 0.0456\footnoteref{foot:contour}      & \emph{0.04640497}  & -           & -                & $-$8.0\footnoteref{foot:contour}       & \emph{$-$15.04}  & 603.3   & -      &                         \\
&XV,\,XVI   & 19648.7\footnoteref{foot:contour}                                                           & ${^2}\Pi-{^2}\Pi$        & 0.0460\footnoteref{foot:contour}      & \emph{0.04640497}  & -           & -                & $-$24.4\footnoteref{foot:contour}      & \emph{$-$15.04}  & 658.9   & -      & $\{6\}_0^1$ stretch                       \\
&XVII & 19926.41\footnoteref{foot:contour}                                                                  & ${^2}\Pi-{^2}\Pi$        & 0.0460\footnoteref{foot:contour}      & \emph{0.04640497}  & -           & -                & $-$14.50\footnoteref{foot:contour}\footnoteref{foot:tentative}               & \emph{$-$15.04}     & 936.6   & -      &                         \\
&XVIII & 20627.2\footnoteref{foot:contour}                                                                  & ${^2}\Pi-{^2}\Pi$        & 0.0461\footnoteref{foot:contour}      & \emph{0.04640497}  & $-$40\footnoteref{foot:contour}    & \emph{$-$7.12}   & $-$20.3\footnoteref{foot:contour}      & \emph{$-$15.04}  & 1637.4  & -      & [\,$\{4\}_0^1$\,] stretch                \\
&XIX  & 21070.6\footnoteref{foot:contour}                   & ${^2}\Pi-{^2}\Pi$        & 0.0454\footnoteref{foot:contour}      & \emph{0.04640497}  & 30\footnoteref{foot:contour}   & \emph{$-$7.12}   & $-$16.8\footnoteref{foot:contour}   & \emph{$-$15.04}  & 2080.8  & -      & $\{3\}_0^1$ stretch                \\
&     &                                                                               &                          &                 &                    &             &                  &                 &                    &         &          &                         \\

$^{13}$C$_6$H&I,\,II\footnote{Data from Ref. [\!\!\citenum{Bacalla2015}\,].} & 18992.116(1)  & ${^2}\Pi-{^2}\Pi$  & 0.042218(17)    & 0.042973(16)       & -           & -                & $-$20.78(13)    & $-$11.62(13)       & 0.0     & 2.3      & $0_0^0$                 \\
&III  & 19066.9716(16)                                                              & ${^2}\Sigma-{^2}\Sigma$  & 0.04227(4)      & 0.04308(4)         & 34(3)       & $-$4(3)          & -               & -                  & 74.9    & 0.1      & $\mu\{11\}_1^1$ bend           \\
&IV   & 19192.1414(17)                                                              & ${^2}\Sigma-{^2}\Sigma$  & 0.042235(3)     & \emph{0.04308}     & 69.72(11)   & \emph{$-$4}      & -               & -                  & 200.0   & $-$5.2   & $\mu\{10\}^1\ \mu\{11\}_1$ bend \\
&V    & 19215.202(2)                                                                & ${^2}\Sigma-{^2}\Sigma$  & 0.042443(5)     & \emph{0.04308}     & 90.22(15)   & \emph{$-$4}      & -               & -                  & 223.1   & $-$5.1   & $\kappa\{10\}^1\ \mu\{11\}_1$ bend \\
&VI   & 19274.22\footnoteref{foot:contour}                                                                  & ${^2}\Sigma-{^2}\Sigma$  & 0.0422\footnoteref{foot:contour}      & \emph{0.04308}     & $-$4\footnoteref{foot:contour}   & \emph{$-$4}      & -               & -                  & 282.1   & $-$9.8   & [\,$\mu \{9\}^1\ \mu\{11\}_1$\,] bend      \\
&VII  & 19288.62\footnoteref{foot:contour}                                                                  & ${^2}\Pi-{^2}\Pi$        & 0.0423\footnoteref{foot:contour}      & \emph{0.04297}     & -           & -                & $-$11.40\footnoteref{foot:contour}\footnoteref{foot:tentative}     & \emph{$-$11.62}       & 296.5   & $-$9.9   &                         \\
&VIII & 19316.45\footnoteref{foot:contour}                                                                  & ${^2}\Pi-{^2}\Pi$        & 0.0425\footnoteref{foot:contour}      & \emph{0.04297}     & -           & -                & -               & -                  & 324.3   & $-$10.0  &       \\
&     &                                                                               &                          &                 &                    &             &                  &                 &                    &         &          &                         \\

C$_6$D&I,\,II\footnoteref{foot:Zhao2011b}  & 19041.2564(5)                          & ${^2}\Pi-{^2}\Pi$        & 0.043592(4)     & \emph{0.04429243}  & $-$6.3(7)   & \emph{$-$3.85}   & $-$24.07(2)  & \emph{$-$15.13}  & 0.0     & 51.5     & $0_0^0$                 \\
&III\footnoteref{foot:Zhao2011b}           & 19115.5251(10)                         & ${^2}\Sigma-{^2}\Sigma$  & 0.0436478(19)   & \emph{0.0443852}   & 39.68(10)   & \emph{$-$0.631}  & -               & -                  & 74.3    & 48.7     & $\mu\{11\}_1^1$ bend           \\
&IV                                        & 19243.221(2)                              & ${^2}\Sigma-{^2}\Sigma$  & 0.043638(5)     & \emph{0.0443852}   & 57.89(11)   & \emph{$-$0.631}  & -               & -                  & 202.0   & 45.9     & $\mu\{10\}^1\ \mu\{11\}_1$ bend \\
&V                                         & 19271.454(5)                           & ${^2}\Sigma-{^2}\Sigma$  & 0.043721(11)    & \emph{0.0443852}   & 64.6(3)     & \emph{$-$0.631}  & -               & -                  & 230.2   & 51.2     & $\kappa\{10\}^1\ \mu\{11\}_1$ bend \\
&VI                                        & 19327.10\footnoteref{foot:contour}                             & ${^2}\Sigma-{^2}\Sigma$  & 0.0429\footnoteref{foot:contour}      & \emph{0.0443852}   & $-$1\footnoteref{foot:contour}       & \emph{$-$0.631}                & -               & -                  & 285.8   & 43.1     & [\,$\mu \{9\}^1\ \mu\{11\}_1$\,] bend      \\
&VII                                       & 19342.16\footnoteref{foot:contour}                             & ${^2}\Pi-{^2}\Pi$        & 0.0435\footnoteref{foot:contour}      & \emph{0.04429243}  & -           & -   & $-$14.68\footnoteref{foot:contour}\footnoteref{foot:tentative}     & \emph{$-$15.13}  & 300.9   & 43.6     &                         \\
\bottomrule \\
\end{tabular}
}
\raggedright \footnotesize{Note: Except for Bands I and II and for Band III, we assume $D' \approx D''$: $1.35\times10^{-9}$~\wn\ (\ce{C6H}) and $1.2\times10^{-9}$~\wn\ (\ce{^{13}C6H} and \ce{C6D}).}
\renewcommand{\mpfootnoterule}{}
\end{sidewaystable*}

\subsection*{Bands I and II: $0^{0}_{0}\ $B$^2\Pi_\Omega - $X$^2\Pi_\Omega$ origin band}
\begin{figure}[]
  \centering
  \includegraphics[width=0.48\textwidth]{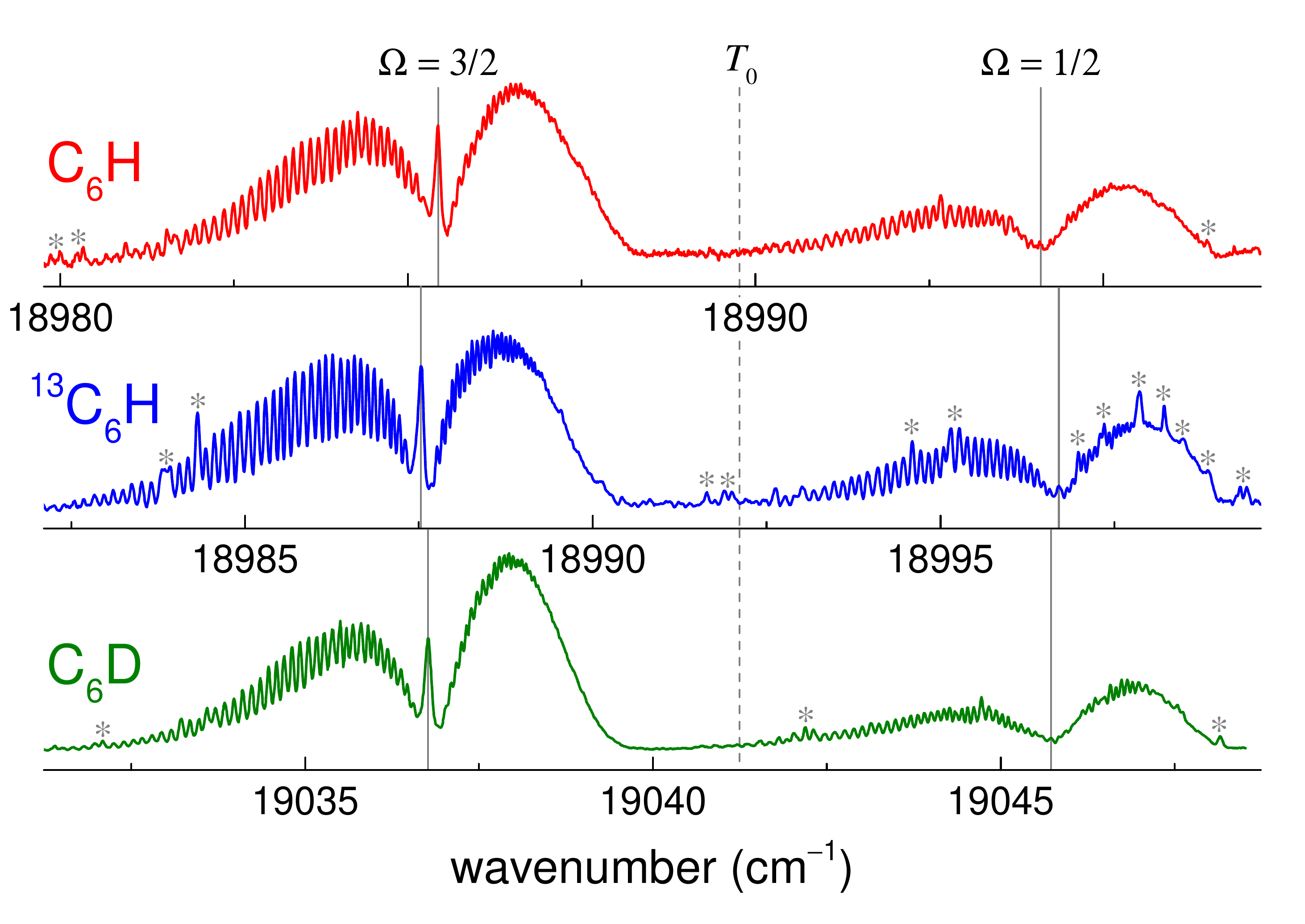}
  \caption{The B$^2\Pi - $X$^2\Pi$ electronic origin band (referred to as Bands I and II) of C$_6$H, measured with a slit plasma nozzle, with the fully $^{13}$C-substituted and the deuterated isotopologue. All three spectra, measured in the slit jet configuration, are plotted here on the same scale with their band origins aligned (denoted by $T_0$). Peaks marked with an asterisk ($*$) are blended transitions from smaller species.}
  \label{fig:compiled-band1and2}
\end{figure}

Fully resolved rovibronic spectra of the B$^2{\Pi}-$X$^2{\Pi}$ origin band transition have been reported before.~\cite{Linnartz1999,Zhao2011b,Bacalla2015} Figure~\ref{fig:compiled-band1and2} shows the cavity ring-down spectra of C$_6$H, $^{13}$C$_6$H, and C$_6$D. The spectra are repeated for completeness and as an introduction to the interpretation of the other $^2\Pi - {^2}\Pi$ bands presented later.

In each spectrum two spin-orbit components of a band system are shown, corresponding to the (0,0) band origin of the B$^2{\Pi}_{3/2}-$X$^2{\Pi}_{3/2}$ (left of $T_0$) and the B$^2{\Pi}_{1/2}-$X$^2{\Pi}_{1/2}$ (right) transition. Each component displays largely rotationally resolved $P$ and $R$ branches. In the B$^2{\Pi}_{3/2}-$X$^2{\Pi}_{3/2}$ case, the $Q$ branch is clearly visible; in the B$^2{\Pi}_{1/2}-$X$^2{\Pi}_{1/2}$ case it is much harder to discriminate. The two subbands are split by the difference in spin-orbit splitting between the ground and the electronically excited state, $A'-A''$. Depending on whether this number is positive or negative, the B$^2{\Pi}_{3/2}-$X$^2{\Pi}_{3/2}$ component is situated to lower or higher energy of the B$^2{\Pi}_{1/2}-$X$^2{\Pi}_{1/2}$ component, assuming that both ground state and excited state have a regular or inverted spin-orbit splitting. As the intensity of the $Q$ branch scales with $\Omega^2$, it is clear that the lower energy component corresponds with the B$^2{\Pi}_{3/2}-$X$^2{\Pi}_{3/2}$ system.~\cite{Linnartz1999} Also, a full rotational analysis (results presented in Table \ref{tab:summary}) shows that the band gap amounts to roughly $10B$, with $B$ as the rotational constant, instead of $6B$ as typical for the B$^2{\Pi}_{1/2}-$X$^2{\Pi}_{1/2}$ band. As a consequence the assignment of the two bands is straightforward and in combination with microwave data \cite{Pearson1988} the spin-orbit splittings are set as an inverted system with $A''=-15.04$~cm$^{-1}$ and $A'=-23.69$ cm$^{-1}$. Based on these numbers and the overall pattern, it is also possible to derive the spin-orbit and rotational temperatures for our plasma settings, yielding 13--22~K.

These findings apply with approximately the same values to all three species. The fitting of the observed spectra is rather straightforward as very precise microwave constants are available to characterize the electronic ground state, but only for the case of C$_6$H and C$_6$D.~\cite{Linnartz1999} Since there are no ground state constants for the $^{13}$C$_6$H species from microwave or infrared spectroscopy, both the upper and the lower energy levels of the B$^2{\Pi}-$X$^2{\Pi}$ (0,0) transition are optimized in the fitting,~\cite{Bacalla2015} and the resulting ground state molecular constants will be used for the simulation and fitting of the other \ce{^{13}C6H} bands.

\begin{figure}[t]
  \centering
  \includegraphics[width=0.48\textwidth]{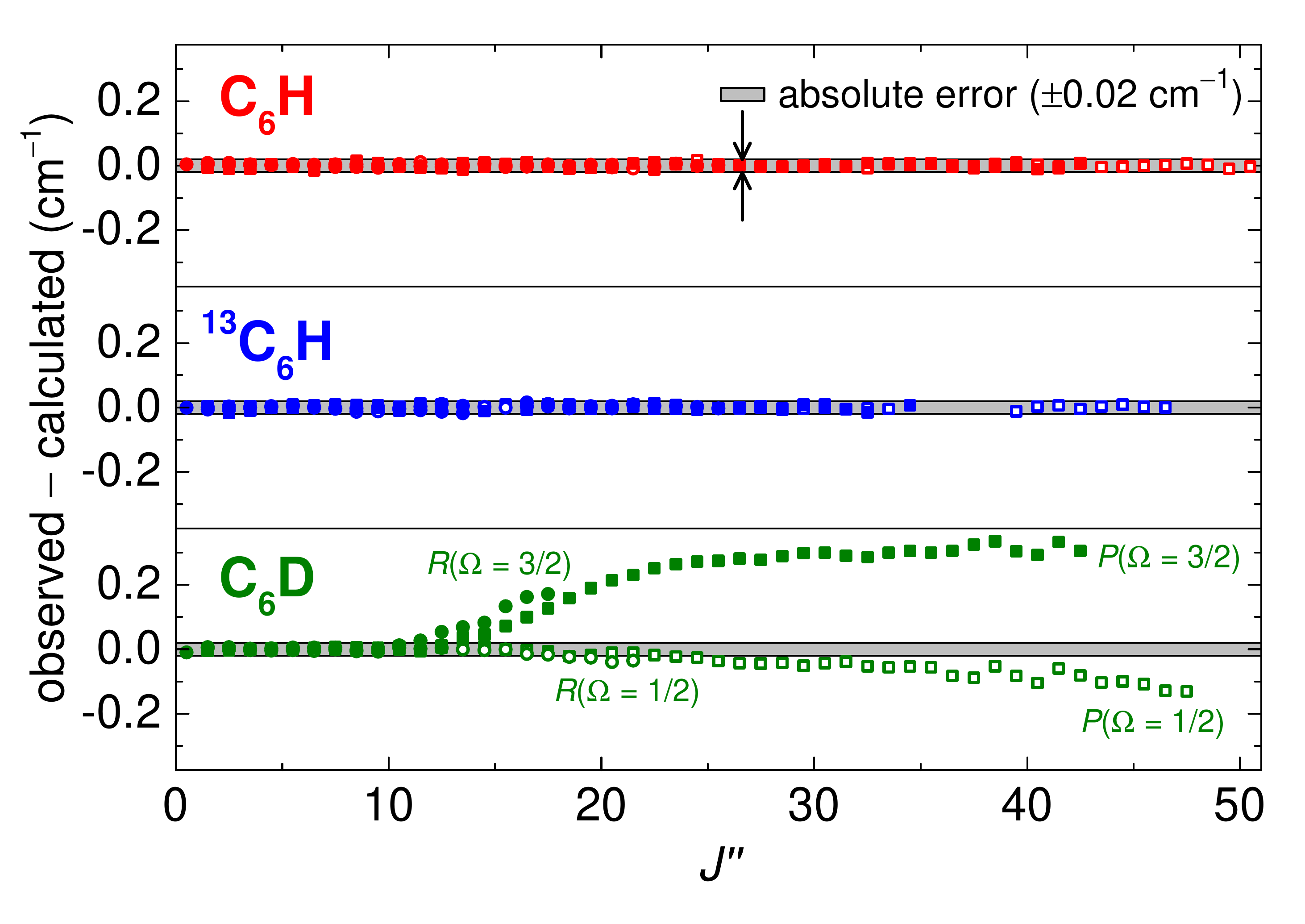}
  \caption{Residuals plotted for each of the observed rotational transitions in the electronic origin band. Squares $(\boxvoid)$ denote $P$ branch lines and circles $(\ovoid)$ to $R$ branch lines, with the filled symbols referring to $\Omega=3/2$ and open symbols to $\Omega=1/2$. Notice that for both C$_6$H and $^{13}$C$_6$H, residuals fall within the uncertainty margins while in C$_6$D they deviate for higher $J''$.}
  \label{fig:Band-IandII-residuals}
\end{figure}

In previous work, extensive lists with line positions have been given for these bands, and using \texttt{PGOPHER} accurate molecular parameters have been derived for \ce{C6H} and \ce{C6D}~\cite{Zhao2011b} as well as for the $^{13}$C-substituted species.~\cite{Bacalla2015} Shown in Fig.~\ref{fig:Band-IandII-residuals} are the differences between the observed and calculated (o--c) values which are derived from these line lists (see \href{dx.doi.org/10.1021/acs.jpca.6b06647}{Supporting Information}). In comparing the behavior of the residuals for the three molecules, the o--c values of \ce{C6D} deviate outside the experimental uncertainty. For $J'' > 15.5\ (10.5)$ in the B$^2{\Pi}_{3/2}-$X$^2{\Pi}_{3/2}$ (B$^2{\Pi}_{1/2}-$X$^2{\Pi}_{1/2}$) component, residuals can become an order of magnitude larger than the absolute uncertainty of $\pm$0.02~cm$^{-1}$. The deviation for the B$^2{\Pi}_{3/2}-$X$^2{\Pi}_{3/2}$ component starts early on and goes to a much larger value as compared to that of the B$^2{\Pi}_{1/2}-$X$^2{\Pi}_{1/2}$ component. This behavior is indicative of a perturbation of the excited state level structure in C$_6$D.

\subsection*{Band III: $\{11\}_1^1$ ${\mu}^2{\Sigma}-{\mu}^2{\Sigma}$ vibronic band}
\begin{figure}[ht!]
  \centering
  \includegraphics[width=0.48\textwidth]{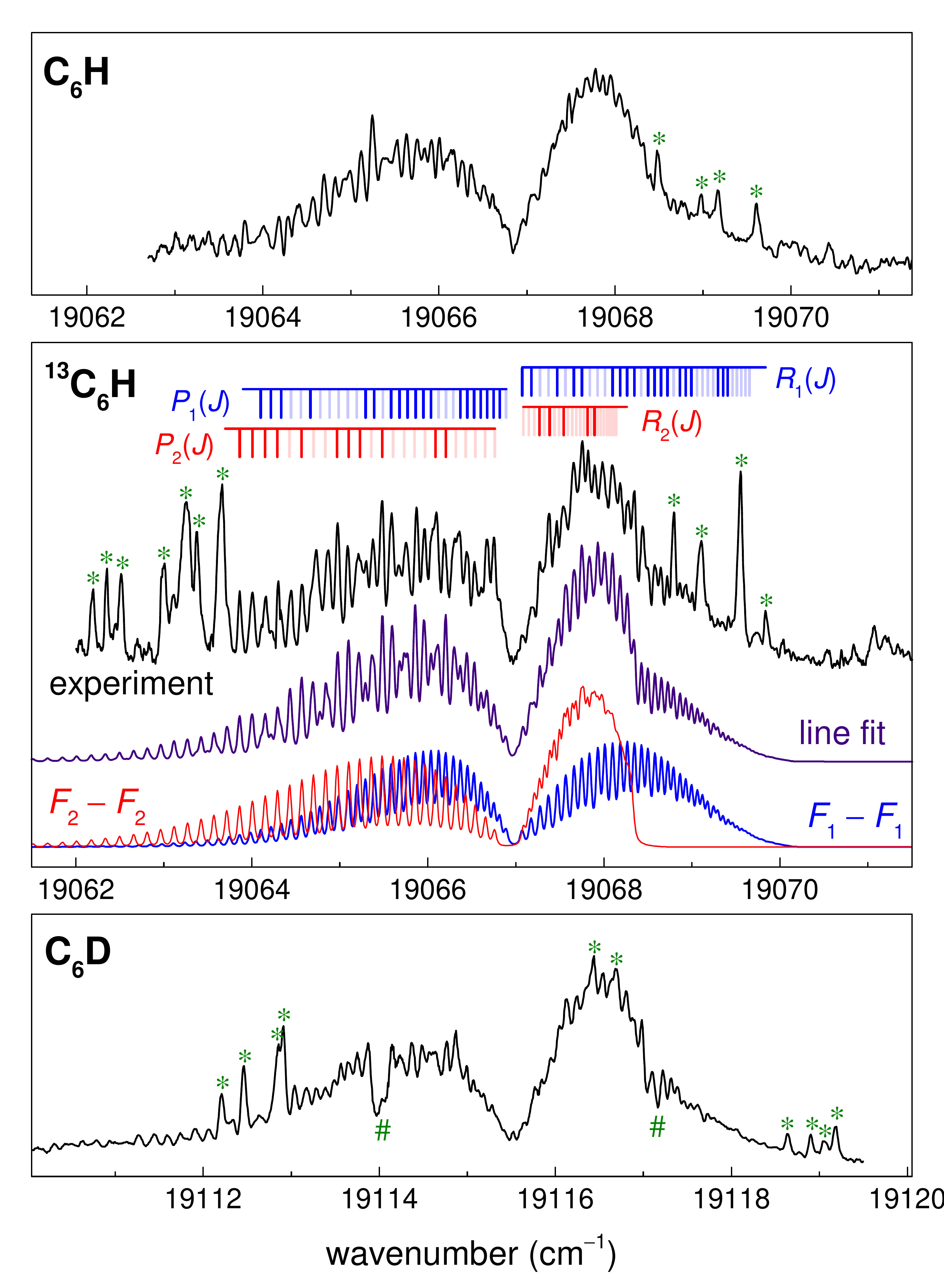}
  \caption{The $\{11\}_1^1$ ${\mu}^2{\Sigma}-{\mu}^2{\Sigma}$ vibronic band (Band III), for all three isotopologues measured in the slit jet configuration. For each branch, the lines are assigned with heavy tick marks. Blended lines from small species are marked with asterisks ($*$). Dips, that are reproduced in independent scans, are present in the C$_6$D spectrum and are marked with hashes~($\#$). Rotational analyses of \ce{C6H} and \ce{C6D} spectra have been presented in Ref. [\!\!\!\citenum{Zhao2011b}\,].}
  \label{fig:Band-III}
\end{figure}

First among the newly recorded vibronic spectra is the $\{11\}_1^1$ ${\mu}^2{\Sigma}-{\mu}^2{\Sigma}$ transition of \ce{^{13}C6H}, highlighted in the middle panel of Fig.~\ref{fig:Band-III}. While this band exhibits clearly discernible $P$ and $R$ branches, no $Q$ branch is observed, indicative of a $\Sigma-\Sigma$ type of transition, which does not allow for a change of electronic angular momentum and thus, $\Delta J = 0$ is forbidden.~\cite{Herzberg1950} For C$_6$H and C$_6$D corresponding bands have been identified as originating from the $\{11\}_1^1$ ${\mu}^2{\Sigma}-{\mu}^2{\Sigma}$ vibronic hot band, and full rotational and Renner-Teller (RT) analyses have been presented previously.~\cite{Zhao2011b} Here we use the information from the rotational analysis of C$_6$H and C$_6$D to fit the $^{13}$C$_6$H line positions. First, a spectral simulation using the C$_6$H constants (after isotopic scaling) is used to assign the individual $^{13}$C$_6$H transitions. These are then included in a fit with \texttt{PGOPHER} using a regular $^2\Sigma-{^2}\Sigma$ Hamiltonian, yielding the $^{13}$C$_6$H constants. The resulting simulation is shown in Fig.~\ref{fig:Band-III}.

Apart from the blended absorption lines from small species, the fit reproduces the experimental spectrum convincingly, including the irregular features in the $P$ branch and the bump in the $R$ branch. We can decompose this fitted curve into two subbands labelled $F_1 - F_1$ and $F_2 - F_2$ to show clearly the doublet $P$ and doublet $R$ band structure. The $F_1 - F_1$ subband gives rise to $P_1(J)$ and $R_1(J)$ branches, and the $F_2 - F_2$ subband to $P_2(J)$ and $R_2(J)$ branches. The overlap of the $P_1(J)$ and $P_2(J)$ branches reproduces the experimentally observed spectral interference, and a more intense $R_2(J)$ compared to the $R_1(J)$ branch produces the step-like phenomenon at $19\,068.3$~\wn. The relative intensity of the two subbands also confirms the vibronic assignment since different values for the rotational constants result in different band contours. It is interesting to note that in this way the somewhat irregular appearance of Band III can be fully reproduced. The derived molecular constants are listed in Table \ref{tab:summary}.

\subsection*{Band IV: $\{10\}^1\{11\}_1\ \mu^2\Sigma - \mu^2\Sigma$ vibronic band}
\begin{figure}[ht!]
  \centering
  \includegraphics[width=0.48\textwidth]{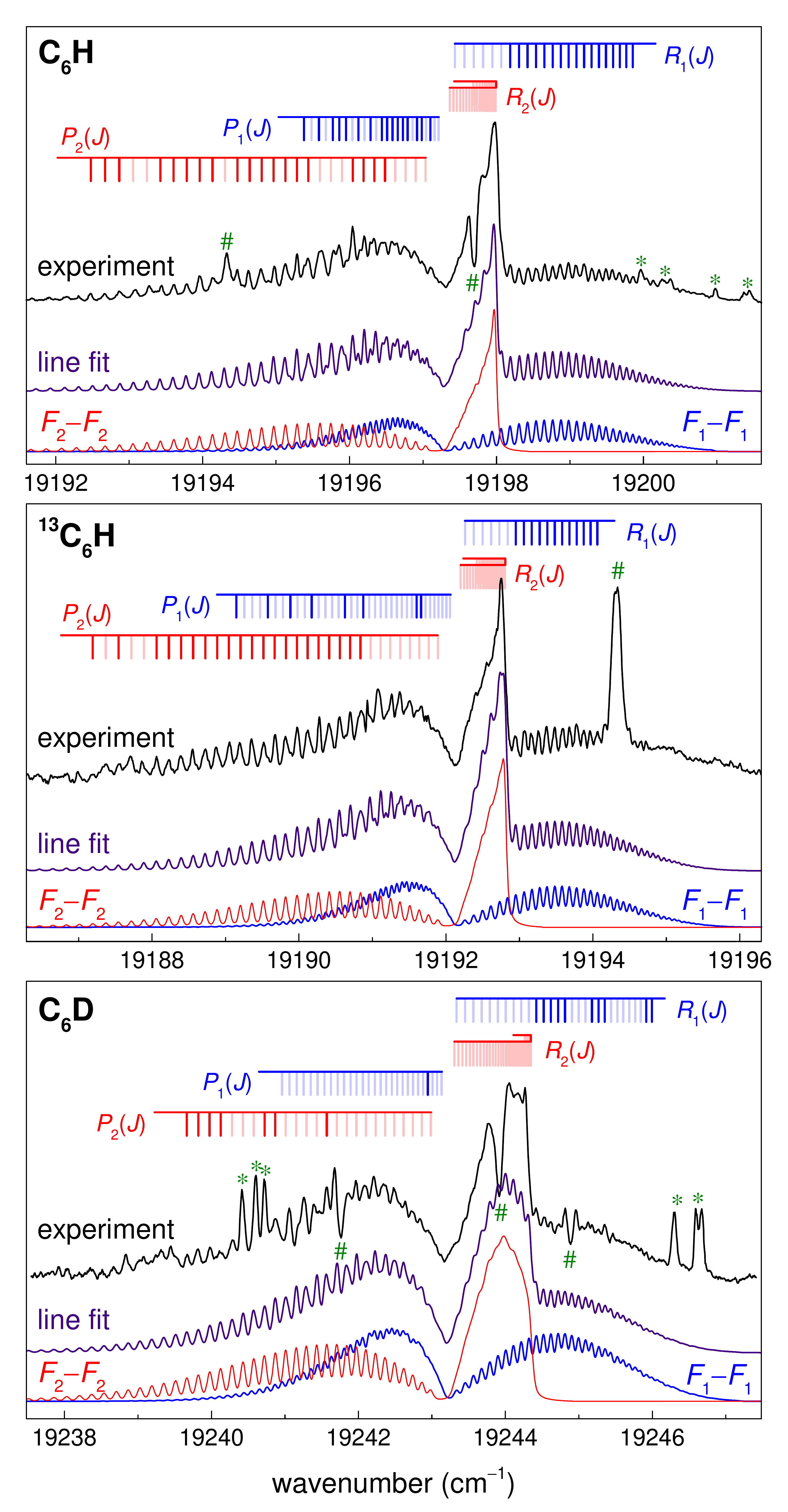}
  \caption{Recording of the $\{10\}^1\{11\}_1\ \mu^2\Sigma - \mu^2\Sigma$ vibronic band (Band IV) for \ce{C6H} (upper), \ce{^{13}C6H} (middle), and \ce{C6D} (lower). Both spectra for \ce{C6H} and \ce{C6D} were measured using the slit nozzle configuration. For \ce{^{13}C6H} the pinhole nozzle configuration was used with an admixture of argon gas to reduce Doppler broadening. Blending absorptions from smaller species are marked with asterisks ($*$). For features marked with a hash~($\#$), see text.}
  \label{fig:Band-IV}
\end{figure}

A well resolved spectrum of this band system (Band IV) is reported for the first time and has been measured for all three isotopologues (Fig.~\ref{fig:Band-IV}). The appearance of the band structure for the three isotopologues is very similar, providing confidence in assigning the three spectra to the same band system. Largely resolved $P$ and $R$ branches are observed at the low and high energy side of an unresolved and structureless feature, more or less at the position where a $Q$ branch might be expected. However, in a simulation it can be shown that these unresolved features are due to the $R_2(J)$ branch of the $F_2 - F_2$ component in a $^2\Sigma - {^2}\Sigma$ band. The intensities of these unresolved features scale with that of the $P$ and $R$ branches and must, therefore, be part of this band system.

The different appearance, compared to Band III, is caused by a larger ratio between the effective spin-rotation constant $\gamma$ and the effective $B$ rotational constant (see Table \ref{tab:summary}), causing the lines in the $R_2(J)$ branch to move closer together, forming a band head which is uncommon for low rotational temperatures. This is illustrated in Fig.~\ref{fig:Band-IV} where the fits of the individual subbands as well as the composite spectrum are shown. Despite these overlapping transitions, it is possible to assign some of the rotational levels and perform a fit. This is because at the low energy side the $P_2(J)$ transitions contribute largely to the intensity, while at the high energy side the $R_1(J)$ branch is the sole contributor to the spectrum. In the experimental spectrum the individual rotational lines are largely resolved and these are well reproduced by the fitted spectrum. This even applies to the overlapping rotational structure in the energy region just at the low energy side of the $R_2(J)$ band head.

The observed spectra are well reproduced using the corresponding ground state constants derived for Band III, confirming that Band IV originates from the same $\{11\}_1\ {\mu}^2\Sigma$ state. Also the relative intensities of the subbands $F_1 - F_1$ and $F_2 - F_2$ further support this assignment. It should be noted, though, that for C$_6$H and C$_6$D, a narrow depletion in the absorption signal has been found in the $R_2(J)$ band head region that is not reproduced in the fit. Should this be a perturbation in the upper rotational states, a corresponding dip should show up in the $P$ branch, but this is not observed. A simulation of the spectra for C$_6$H and C$_6$D including these anomalies, marked by a hash ($\#$), turned out to be impossible. In the same way for $^{13}$C$_6$H a strong and broad line shows up around 19194.3 cm$^{-1}$ with unclear origin. Incidentally the strong line in $^{13}$C$_6$H coincides with a small peak in the \ce{C6H} spectrum, at exactly the same frequency, both marked by a hash ($\#$), again indicative of the same feature of unknown origin; an assignment is not possible at this moment. The molecular constants for this band are collected in Table \ref{tab:summary}.

\subsection*{Band V: $\{10\}^1\{11\}_1\ \kappa^2\Sigma - \mu^2\Sigma$ vibronic band}
\begin{figure}[ht!]
  \centering
  \includegraphics[width=0.48\textwidth]{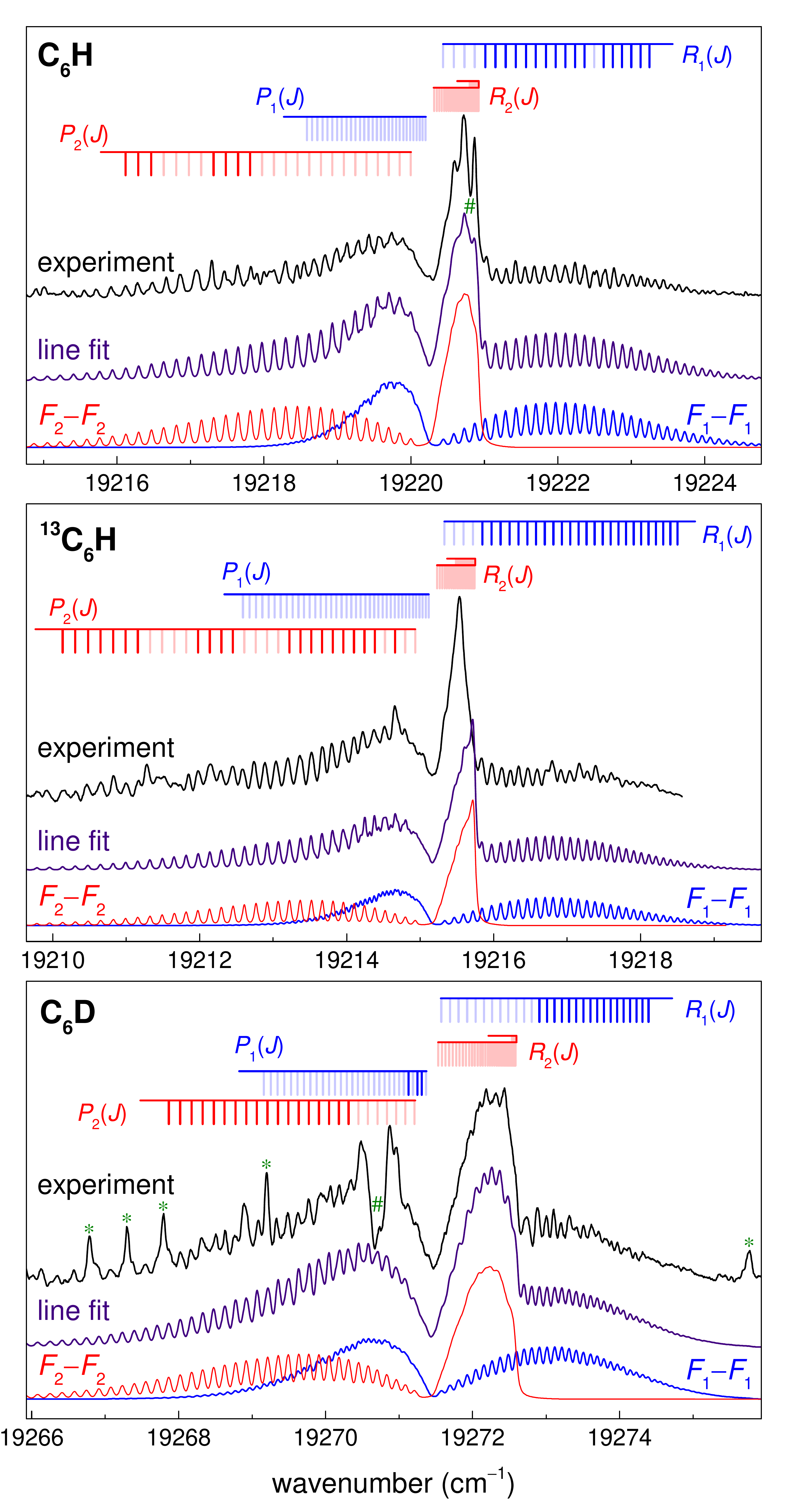}
  \caption{Band V assigned as the $\{10\}^1\{11\}_1\ \kappa^2\Sigma - \mu^2\Sigma$ transition. Again, distinct dips (marked with $\#$) show up in the spectra which cannot be accounted for in the rotational line fit.}
  \label{fig:Band-V}
\end{figure}

Band V (Fig.~\ref{fig:Band-V}) is another well resolved spectrum with a band contour that is very much alike Band IV which strongly suggests that a similar type of electronic transition ($^2\Sigma - ^2\Sigma$) is taking place. With a number of resolved transitions, a rotational line fit can be performed, yielding the molecular constants as listed in Table \ref{tab:summary}. The simulations reproduce the lines and the overall contour, although for \ce{^{13}C6H} the peak of the central spectral feature is slightly off by $\sim$~0.2~cm$^{-1}$ to higher energy. The vibrational assignment of Band V in connection with Band IV is described in the Discussion section.

\subsection*{Band VI: $\mu^2\Sigma - \mu^2\Sigma$ vibronic band}
\begin{figure}[ht!]
  \centering
  \includegraphics[width=0.48\textwidth]{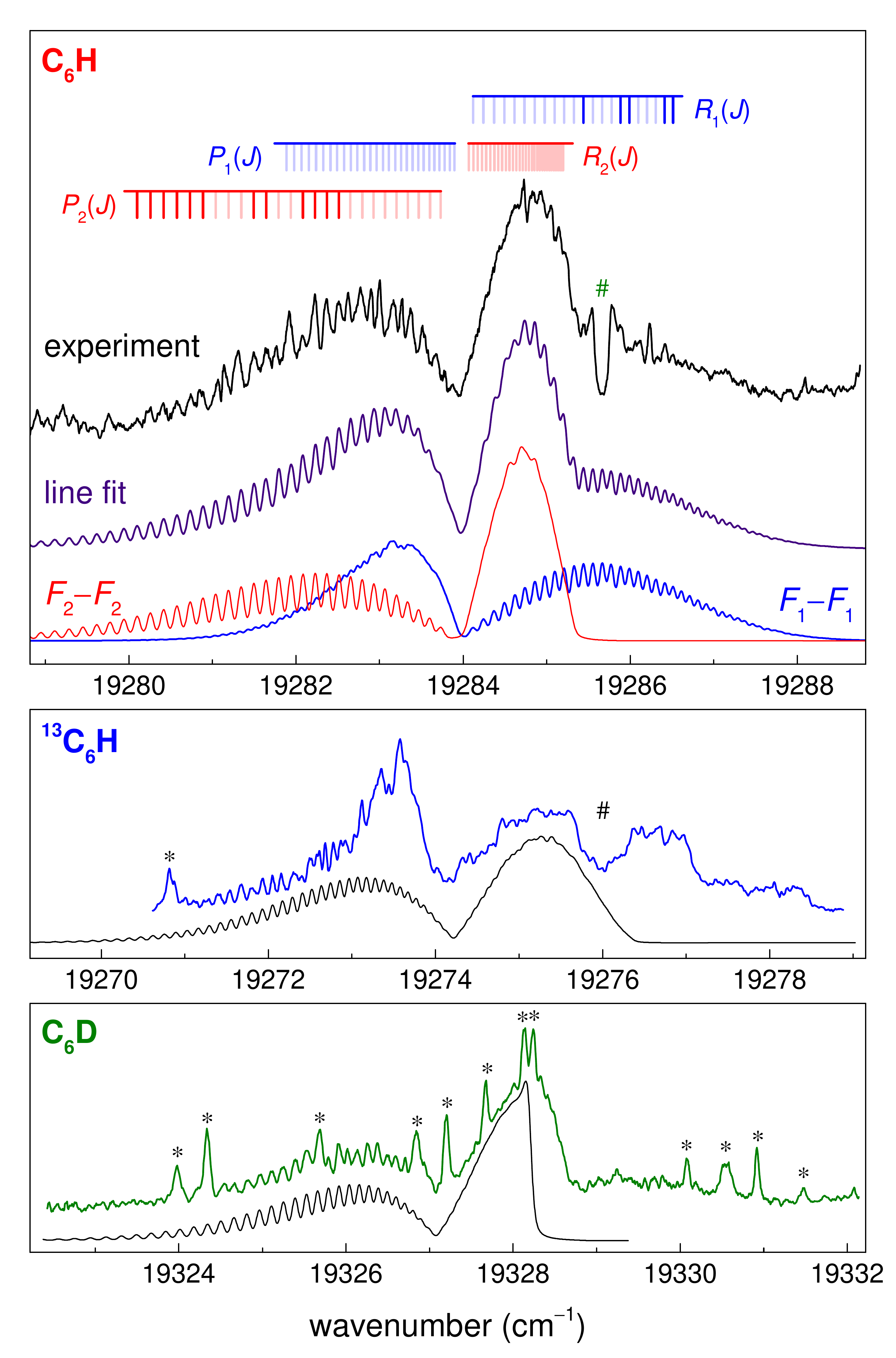}
  \caption{Band VI assigned as a $^2\Sigma - {^2}\Sigma$ transition. All three spectra were recorded using the slit nozzle configuration. Shown in the upper panel is a rotational line fit done for \ce{C6H} with the two band components. For \ce{^{13}C6H} and \ce{C6D} a band contour simulation (in black) was performed. Features denoted by asterisks ($*$) and hashes ($\#$) are discussed in the main text.}
  \label{fig:Band-VI}
\end{figure}

The spectra assigned to Band VI and recorded in the slit jet configuration are presented in Fig.~\ref{fig:Band-VI}. At first glance the contour of Band VI is reminiscent of Band III, rather clearly for \ce{C6H} and \ce{C6D}, pointing to a ${^2}\Sigma - {^2}\Sigma$ transition. For \ce{C6H}, having the best S/N for this band, a fit to the rotational structure is well possible using the ground state constants obtained for the Renner-Teller $\mu^2\Sigma$ ground state. Again the intensity in the $R$~branch is most pronounced and also the intensity drop at $19\,285.4$~\wn\ can be understood as a result of the contribution by the $F_1 - F_1$ and $F_2 - F_2$ subbands. The spectrum for \ce{C6H} has a single artifact feature denoted by a hash ($\#$), while the spectra of \ce{C6D} and \ce{^{13}C6H} are overlaid by multiple artifact features and extra lines associated with small radicals, denoted by asterisks ($*$).

Given the low S/N, for the \ce{^{13}C6H} and the \ce{C6D} spectrum a good line fit is not possible for molecular constants isotopically scaled from the main \ce{C6H} isotopologue. Moreover, anomalous band features make it harder to identify and assign rotational lines. The lack of individually resolved rotational transitions necessitates a contour simulation using the known ground state values and tuning the excited state parameters together with the rotational temperature until the overall shape of the band is matched. Clearly, results obtained using this procedure will not be as precise as the results from a line fit. So as an additional criterion to verify the correct simulation of the spectrum, the spacing between individual (but unassignable) rotational lines has been matched, typically in the $P$~branch.

The spectrum for \ce{C6D} is overlaid by a number of sharp lines (marked by $*$). If this is taken into account, the contour fit yields a reasonable simulation of the observed spectrum, although the sharp drop in the simulated intensity of the $R$~branch is not clearly reproduced.

Of the three spectra, the \ce{^{13}C6H} spectrum exhibits the most severe anomalies which makes it less straightforward to obtain a proper band contour simulation. The increased intensity in the $P$~branch near $19\,273.5$~\wn\ might be associated by the spectral lines of small radicals contributing to the intensity, although no sharp and distinct lines are visible. In the $R$~branch of \ce{^{13}C6H} a double broad feature appears which could be due to an artifact dip (marked by $\#$) or it may well be that the second feature is caused by a different species since it is rather outside the simulated intensity pattern. In any case the observed spectrum deviates quite distinctively from a constructed band contour as shown in the figure, and so the resulting constants should be treated with care.

The molecular constants obtained from a fit of the \ce{C6H} spectrum and a band contour simulation for \ce{C6D} and \ce{^{13}C6H} are listed in Table~\ref{tab:summary}.

\subsection*{Band VII: B${^2}\Pi_\Omega-$X${^2}\Pi_\Omega$ vibronic band}
\begin{figure}[ht!]
  \centering
  \includegraphics[width=0.48\textwidth]{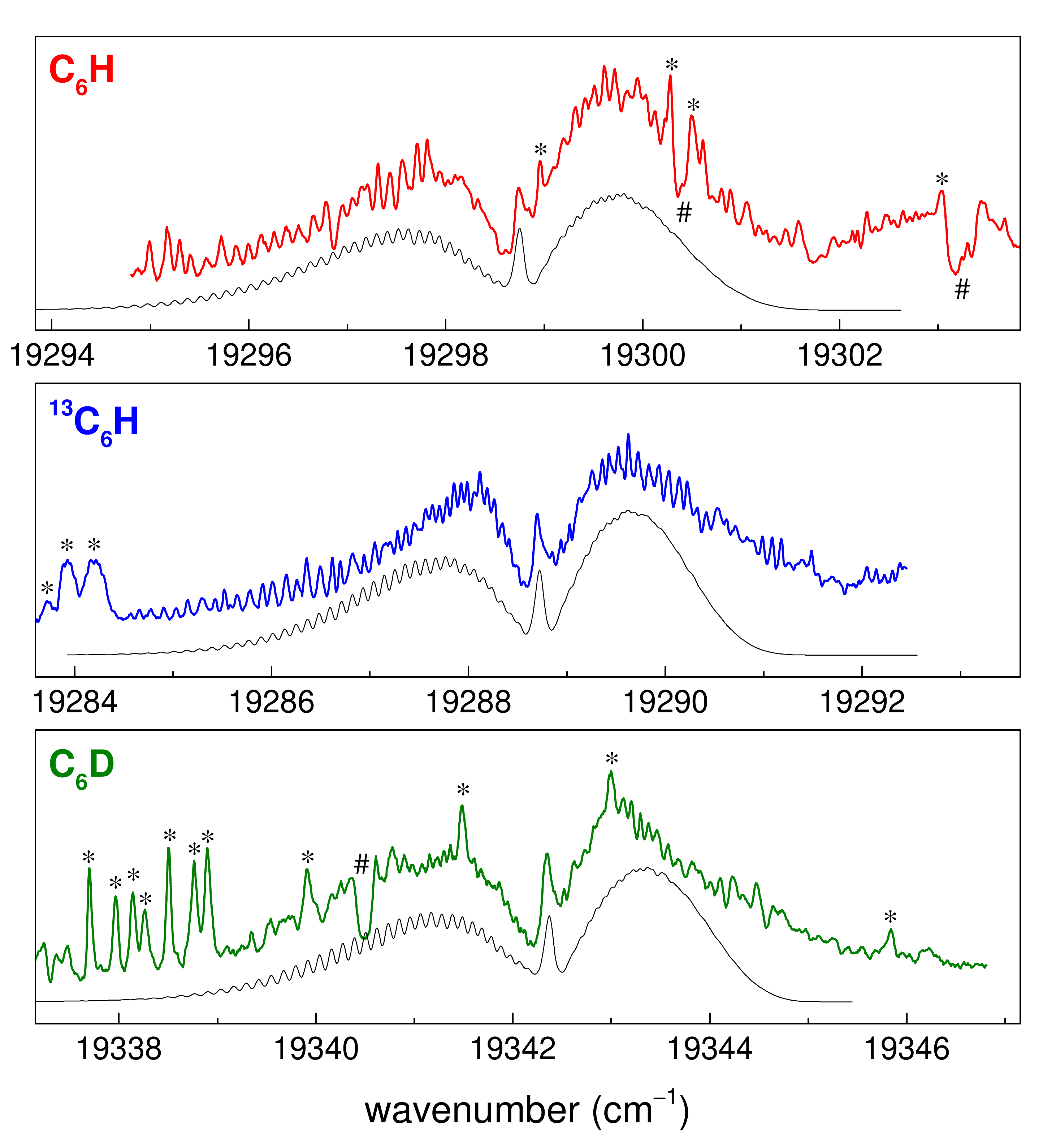}
  \caption{Spectral recordings of Band VII, here identified as a ${^2}\Pi-{^2}\Pi$ transition. All spectra were recorded using the slit nozzle configuration. Again blended lines ($*$) and dips ($\#$) are visible in the spectrum. Simulated band contours are shown in black.}
  \label{fig:Band-VII}
\end{figure}

Spectra of Band VII, recorded in the slit jet configuration for all three isotopologues, are displayed in Fig.~\ref{fig:Band-VII}. The overall shape of the band structures for the three isotopologues is similar, which forms the basis for assigning these spectral features to the same band. In the band gap region between the $P$ and $R$~branches a clear peak reminiscent of a typical $Q$~branch is found. In the spectra for \ce{C6H} and \ce{C6D} some additional lines associated with small radicals (marked with $*$) are visible as well, and their shape is somewhat similar to the $Q$~branch feature. However, because these extra lines do not occur in the spectrum for \ce{^{13}C6H} and the location of the $Q$~branch is in all cases at the appropriate frequency, the identification of a $Q$~branch feature is unambiguous. Hence Band VII is assigned as ${^2}\Pi-{^2}\Pi$.

As was discussed in previous studies of the ${^2}\Pi-{^2}\Pi$ band origin transition in \ce{C6H},~\cite{Linnartz1999} the $Q$~branch is most visible for the $\Omega = 3/2$ spin-orbit component, while it is hardly visible for $\Omega = 1/2$; this intensity relationship is exemplified in the origin band (Fig.~\ref{fig:compiled-band1and2}). Detailed inspection of the recorded $Q$~branches of Band VII for all isotopologues shows that all three are somewhat blue-degraded, unlike in the case of the origin band, where the $Q$~branches are red-degraded. Such a line shape is difficult to explain. A possible rationale is that the two spin-orbit components of the transition, hence ${^2}\Pi_{3/2}-{^2}\Pi_{3/2}$ and ${^2}\Pi_{1/2}-{^2}\Pi_{1/2}$, are overlapped with each other to some extent, which further implies that for this transition, $A' \approx A''$. A similar situation applies to other species, like \ce{HC6H+} and \ce{HC8H+}, where transitions starting from both spin-orbit components coincide and differences are only visible for higher rotational temperatures.~\cite{Pfluger2000} In principle, it is also possible that $A''$ increases, and for a similar spin-orbit temperature, this means that the population in the $\Omega = 1/2$ state decreases. This is not expected, as the ground state splitting should be the same for all $^2\Pi-{^2}\Pi$ transitions studied here. The constructed contours, shown in black in Fig.~\ref{fig:Band-VII}, are based on this assumption. In all three cases the observed spectra are reasonably well reproduced in the contour fits and the underlying molecular constants are listed in Table~\ref{tab:summary}.

\subsection*{Band VIII: B${^2}\Pi_{3/2}-$X$^2\Pi_{3/2}$ vibronic band}
\begin{figure}[ht!]
  \centering
  \includegraphics[width=0.48\textwidth]{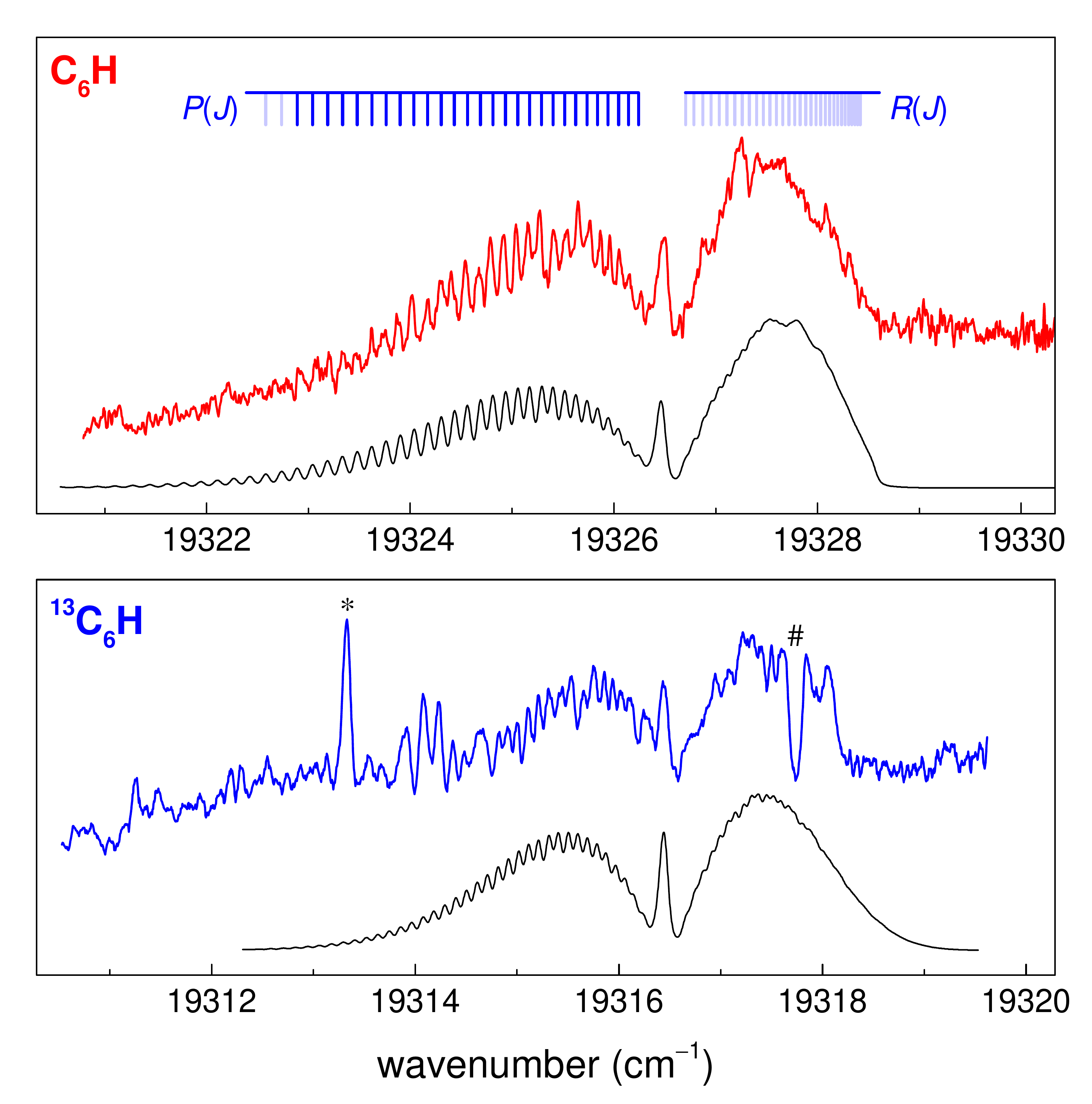}
  \caption{Spectra for Band VIII, assigned as the $\{9\}_0^1\ $B${^2}\Pi_{3/2}-$X$^2\Pi_{3/2}$ transition,  recorded for \ce{C6H} in the slit nozzle configuration and for \ce{^{13}C6H} in the pinhole nozzle configuration. Simulated band contours are shown in black.}
  \label{fig:Band-VIII}
\end{figure}

Figure~\ref{fig:Band-VIII} displays the spectra of Band VIII recorded for \ce{C6H} and \ce{^{13}C6H}. Band VIII was not recorded for \ce{C6D} since in the energy region beyond $\sim$~$19\,350$~\wn, where this band is expected (see also Fig.~\ref{fig:compiled-allbands}), the spectra are covered by very strong absorptions from the \ce{C2} Swan bands.

Again, as in the case of Band VII, the spectra are characterized by the presence of a pronounced $Q$ branch, the reason for assigning this band to a ${^2}\Pi-{^2}\Pi$ type transition. This time, the $Q$ branch lies more to the $P$ branch side, and the line shape of the $Q$~branch is red-degraded, which is an indicator for a ${^2}\Pi_{3/2}-{^2}\Pi_{3/2}$ transition. Indeed both spectra for \ce{C6H} and \ce{^{13}C6H} can be well reproduced by a contour simulation. Values obtained from the contour simulation are listed in Table~\ref{tab:summary}. 

\subsection*{Bands IX and X: B${^2}\Pi_\Omega -$X${^2}\Pi_\Omega$ vibronic band}
\begin{figure}[ht!]
  \centering
  \includegraphics[width=0.48\textwidth]{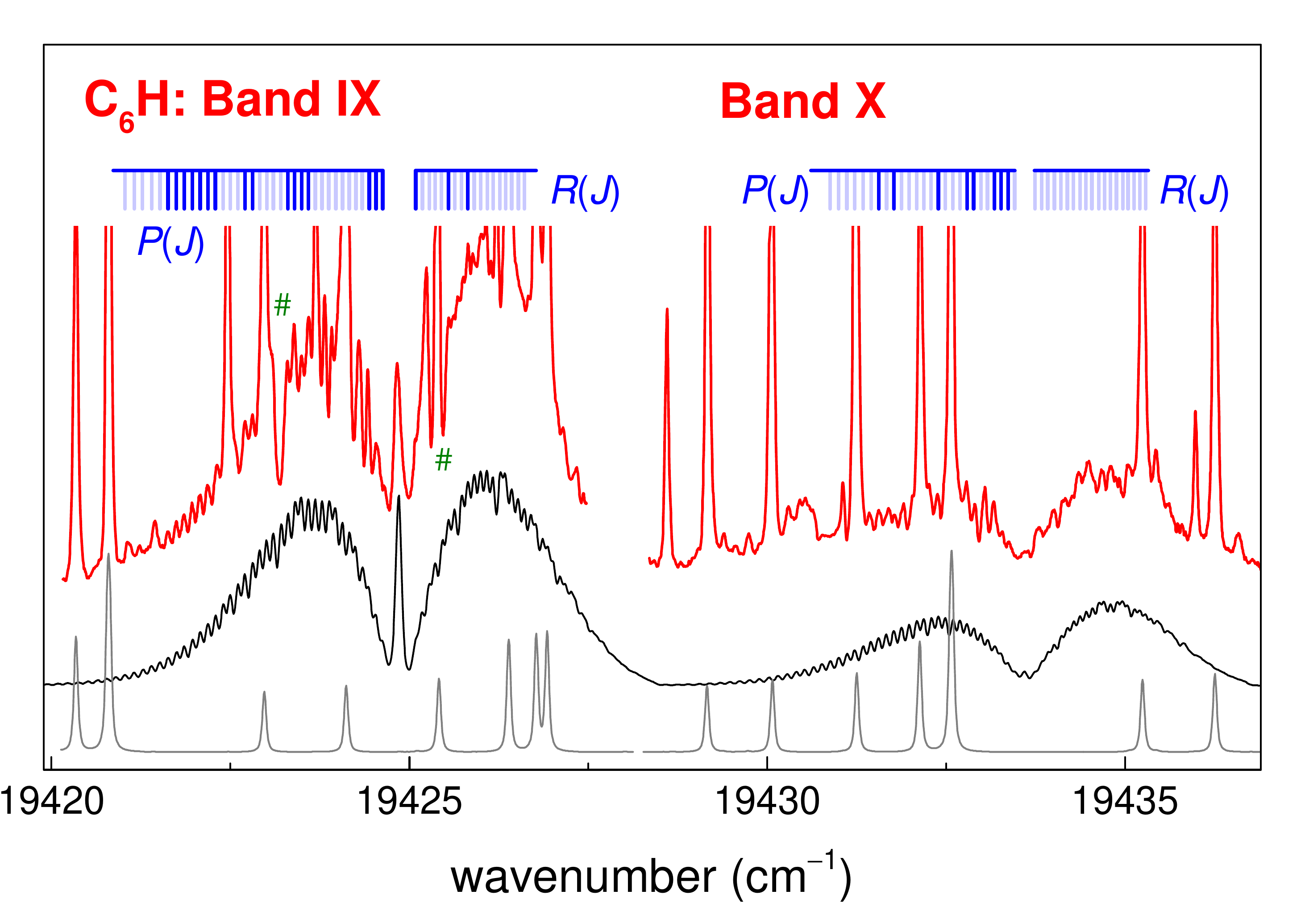}
  \caption{Bands IX and X of \ce{C6H} measured in a slit jet plasma configuration and a rotational line fit for a $^2\Pi_{\Omega} - {^2}\Pi_{\Omega}$ band for both $\Omega=3/2$ and $1/2$ components. At the bottom of the figure a simulated spectrum of overlapping \ce{C2} Swan transitions is plotted.}
  \label{fig:Band-IX-X}
\end{figure}

In the narrow frequency window at $19\,420-19\,440$~\wn\, two more absorption bands are found that are heavily overlaid by strong resonances from \ce{C2}. Using molecular constants from the literature,~\cite{Lloyd1999} these blending peaks can be attributed to the (0,0) origin band of the Swan system of \ce{C2}. This simulation and together with the spectra recorded in the slit jet configuration are displayed in Fig.~\ref{fig:Band-IX-X}. The observed bands (Bands IX and X) have the shape of a ${^2}\Pi_\Omega - {^2}\Pi_\Omega$ pair: very similar to bands I and II. Band IX has a narrow resonance at $19\,424.9$~\wn, which is not reproduced by a Swan band line like most of the other sharp resonances observed. This resonance, identified as a $Q$~branch, is centered in between apparent $P$ and $R$~branch features. Following the argument, as for Bands I and II, that the $\Omega=3/2$ component exhibits a much stronger $Q$~branch than the $\Omega=1/2$ component, a contour fit can be produced for the combined ${^2}\Pi_\Omega - {^2}\Pi_\Omega$ bands using the ground state molecular constants.
The agreement of the excited state spin-orbit constant $A'$ with that of the band origin
provides confidence for a positive assignment of these bands. It is noted that these bands were presented previously in a PhD-thesis by~\citet{Denisov2006} employing cavity ring-down spectroscopy, where the bands were also assigned as a ${^2}\Pi_\Omega - {^2}\Pi_\Omega$ pair. Similarly as in the previously presented vibronic bands, dips show up in the spectra which cannot be reproduced in a fit.

\subsection*{Band XI: B${^2}\Pi_{3/2} - $X${^2}\Pi_{3/2}$ vibronic band}
\begin{figure}[ht!]
  \centering
  \includegraphics[width=0.48\textwidth]{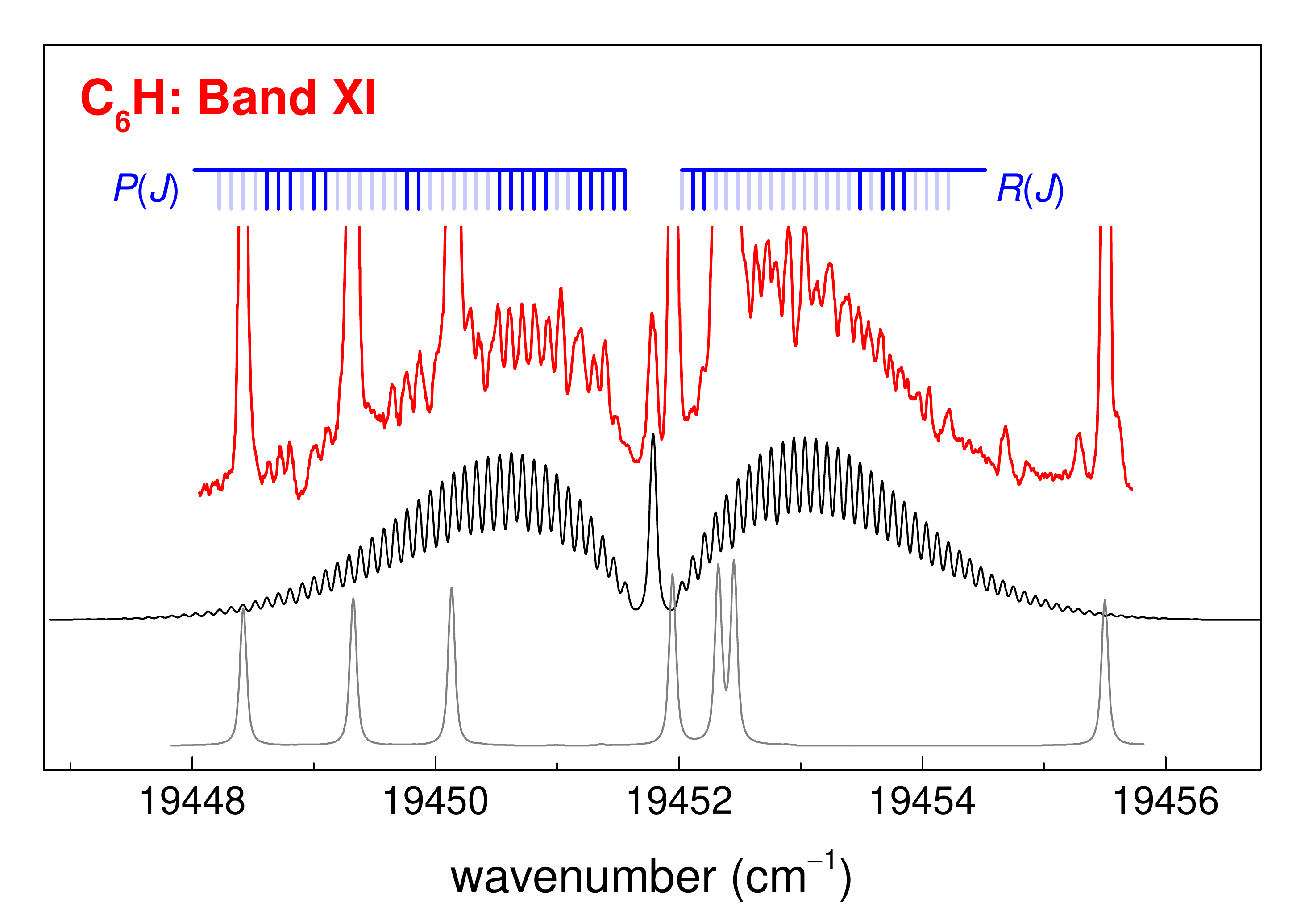}
  \caption{Recording of Band XI of \ce{C6H} in a slit jet plasma configuration and a rotational line fit assuming a ${^2}\Pi_{3/2} - {^2}\Pi_{3/2}$ transition.}
  \label{fig:Band-XI}
\end{figure}

A recording of Band~XI, presented in Fig.~\ref{fig:Band-XI}, is assigned as ${^2}\Pi_{3/2} - {^2}\Pi_{3/2}$ due to the presence of a $Q$~branch. Again the spectrum in this energy range is overlaid by strong resonances due to \ce{C2} Swan (0,0) band absorption, but the prominent absorption feature at $19\,451.8$~\wn\ does not coincide with a predicted \ce{C2} line, and is therefore identified as the $Q$ branch. A number of well-resolved rotational lines in the $P$ and $R$~branches allows for a fit yielding a rotational constant that is indicative of \ce{C6H} (see Table \ref{tab:summary}). The associated ${^2}\Pi_{1/2} - {^2}\Pi_{1/2}$ spin-orbit component could not be observed.

\subsection*{Band XII: ${^2}\Sigma-{^2}\Sigma$ vibronic band}
\begin{figure}[ht!]
  \centering
  \includegraphics[width=0.48\textwidth]{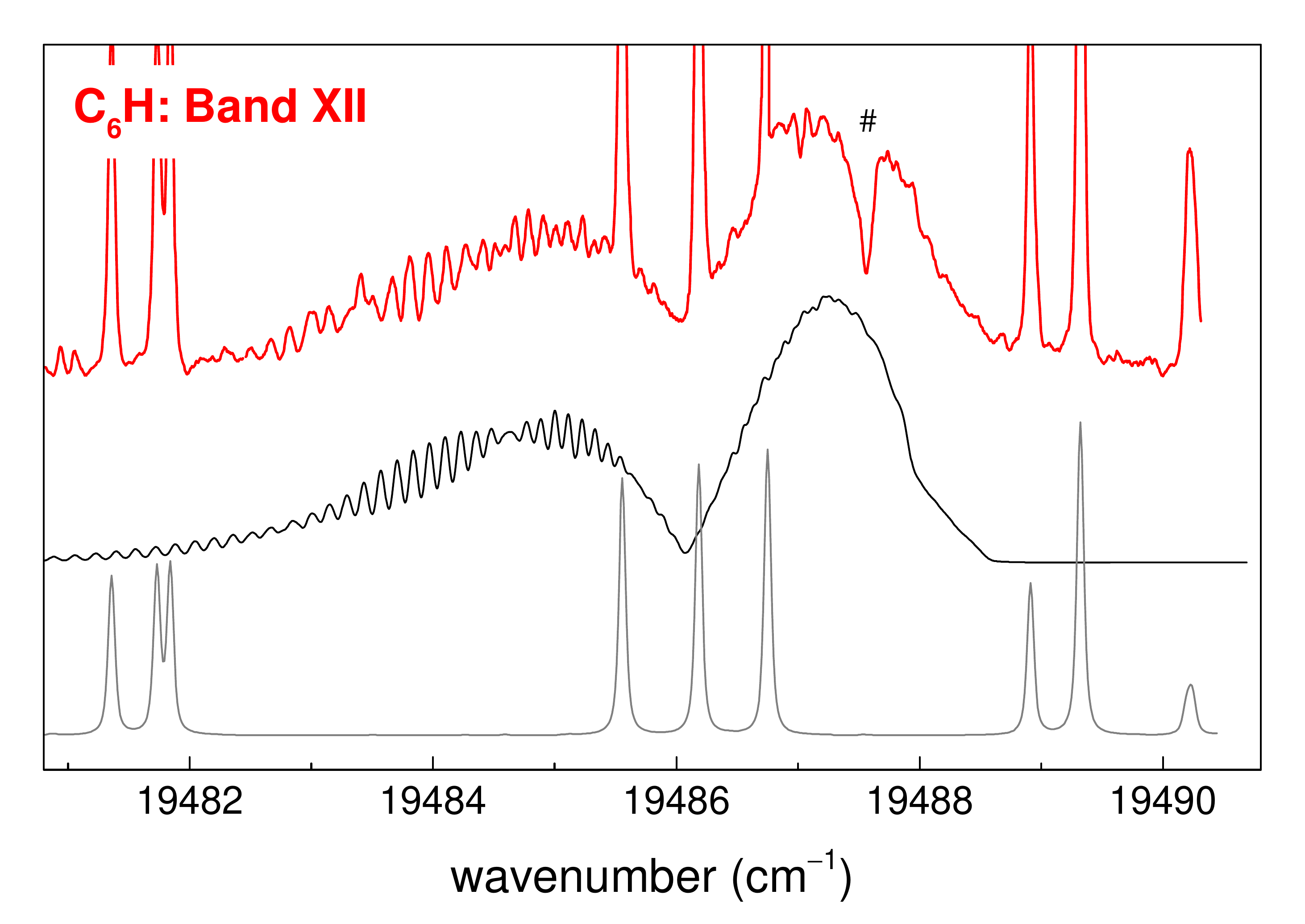}
  \caption{Bands XII of \ce{C6H} measured in the slit plasma nozzle configuration. A simulated contour spectrum of a $^2\Sigma - {^2}\Sigma$ transition is plotted in black. The sharp absorption features can be identified as \ce{C2} Swan band lines for bands (0,0) and (1,1) bands (in gray).}
  \label{fig:Band-XII}
\end{figure}

Figure~\ref{fig:Band-XII} shows a recording of a band centered at $19\,486$~\wn. In a previous study this band feature was associated to the $^2A'' -$X$^2A''$ electronic transition of the \ce{C4H4+} cation,~\cite{Araki2004} while in another study, it was identified as a $^3\Sigma^- -$X$^3\Sigma^-$ transition of the \ce{C6H+} cation.~\cite{Raghunandan2010} Because the spectrum is overlaid by a number of strong and sharp \ce{C2} Swan transitions, particularly in the region where a $Q$~branch of a possible ${^2}\Pi_{3/2} - {^2}\Pi_{3/2}$ transition of \ce{C6H} may be expected, an unambiguous assignment of this band is difficult. Several band contours were simulated, and by imposing a transition of ${^2}\Sigma-{^2}\Sigma$ symmetry and keeping the ground state fixed at values for the $\mu^2\Sigma$, Band XII can be well reproduced with excited state constants
matching those of \ce{C6H}. The fact that the observed interference between simulated $P$ branches for both $F_1 - F_1$ and $F_2 - F_2$ subbands at $\sim$~$19\,484.6$~\wn\ is reproduced in the simulated spectrum lends credit to the assignment of this band to a $^2\Sigma - {^2}\Sigma$ feature and to assign the \ce{C6H} molecule as its carrier. It is interesting to note that the distinctive dip in the $R$~branch at $19\,487.6$~\wn\ in Fig.~\ref{fig:Band-XII}, indicated by $\#$, was also observed in the aforementioned studies.~\cite{Araki2004,Raghunandan2010}  This again demonstrates that these features of decreased absorption are not just artifact features.

\subsection*{Bands XIII and XIV: B${^2}\Pi_\Omega - $X${^2}\Pi_\Omega$ vibronic band}
\begin{figure}[ht!]
  \centering
  \includegraphics[width=0.48\textwidth]{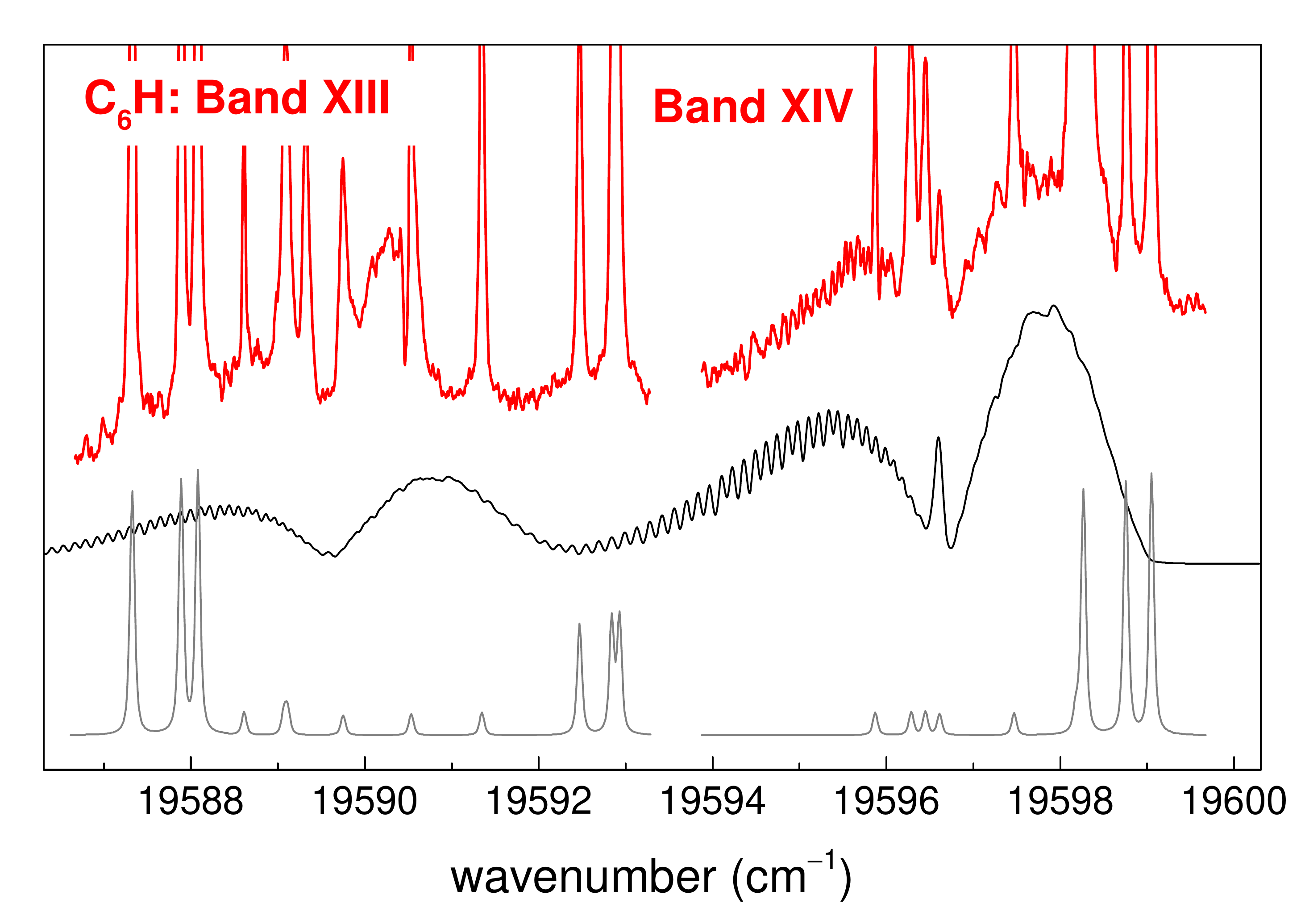}
  \caption{Bands XIII and XIV of \ce{C6H} measured in the slit plasma jet configuration with a simulated spectrum based on a ${^2}\Pi_\Omega - {^2}\Pi_\Omega$ pair with the $\Omega=1/2$ component at lower energies. The sharp and intense absorption peaks arise from \ce{C2} $(0,0)$ and $(1,1)$ Swan bands (in gray).}
  \label{fig:Band-XIII-XIV}
\end{figure}

In the frequency range $19\,587 - 19\,600$~\wn\ another composite band feature has been observed which again is heavily overlaid by sharp \ce{C2} (0,0) and (1,1) Swan band transitions. The spectrum (Bands XIII and XIV), presented in Fig.~\ref{fig:Band-XIII-XIV}, displays the appearance of a ${^2}\Pi_\Omega - {^2}\Pi_\Omega$ pair. Like for the other ${^2}\Pi_\Omega - {^2}\Pi_\Omega$ pairs the $\Omega=3/2$ component (Band XIV) exhibits an identifiable $Q$~branch visible at $\sim 19\,596.5$~\wn, while the other subband feature (Band XIII) does not display such a pronounced $Q$~branch. However, at this location a weak C$_2$ line is predicted in the simulation, so the assignment of the \ce{C6H} bands remains tentative.  In addition the order of the spin-orbit components in the putative $^2\Pi$-doublet is reversed, i.e., $\Omega = 1/2$ falls at lower energies than the $\Omega = 3/2$ component. The molecular constants used for producing the simulated spectrum for the corresponding \ce{C6H} features are listed in Table~\ref{tab:summary}.

\subsection*{Bands XV and XVI: $\{6\}_0^1\ {^2}\Pi_\Omega-{^2}\Pi_\Omega$ vibronic band}
\begin{figure}[ht!]
  \centering
  \includegraphics[width=0.48\textwidth]{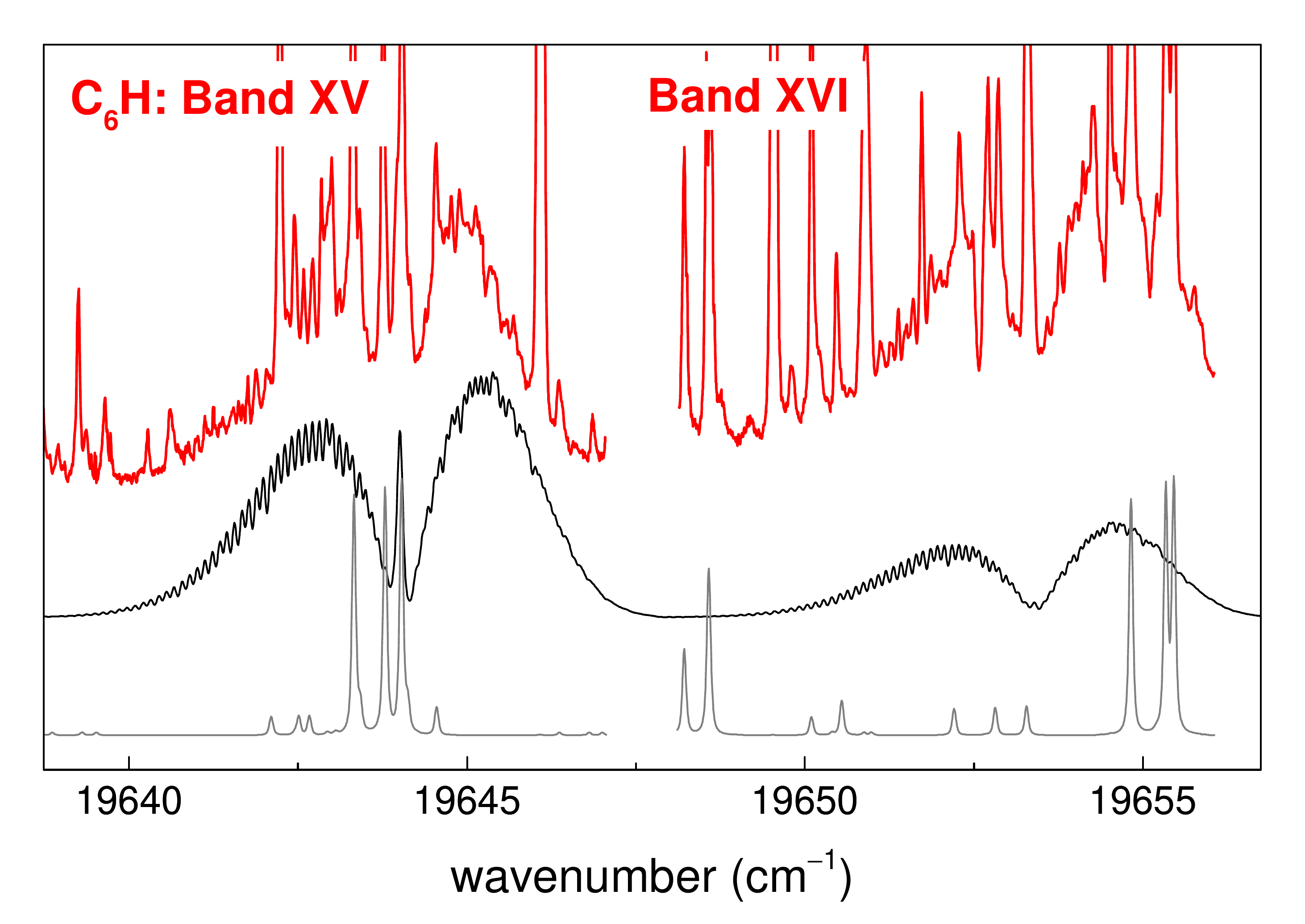}
  \caption{Bands XV and XVI of \ce{C6H} measured with a slit plasma nozzle. The blended absorption peaks arise from \ce{C2} $(0,0)$ and $(1,1)$ Swan bands (in gray).}
  \label{fig:Band-XV-XVI}
\end{figure}

In the frequency range $19\,638 - 19\,656$~\wn\ two sequential bands have been observed for \ce{C6H} in the slit jet discharge expansion. The spectrum, displayed in Fig.~\ref{fig:Band-XV-XVI}, is of low signal-to-noise and overlaid with pronounced sharp resonances due to C$_2$ Swan band lines. By itself the identification of the features remains somewhat unclear, in particular since the intensity of the ${^2}\Pi_{1/2}-{^2}\Pi_{1/2}$ component exhibits about the same intensity as the ${^2}\Pi_{3/2}-{^2}\Pi_{3/2}$ component. Additional measurements using the pinhole configuration and detecting these bands downstream the plasma expansion show that the intensity of the ${^2}\Pi_{1/2}-{^2}\Pi_{1/2}$ component did not drop in a similar fashion as that of the ${^2}\Pi_{1/2}-{^2}\Pi_{1/2}$ component of the origin band, as shown in Ref. [\!\!\citenum{Zhao2011b}\,]. The same features were also observed in the hollow cathode discharge experiment \cite{Kotterer1997} and there they were assigned to a $\{6\}_0^1\ {^2}\Pi_\Omega-{^2}\Pi_\Omega$ transition, based on a comparison with matrix data. From a contour simulation of the spectrum it follows that both subbands, pertaining to $\Omega=3/2$ and $\Omega=1/2$ components, can be well reproduced.
This provides support for a definitive assignment of Bands XV and XVI to $6_0^1\ {^2}\Pi_\Omega-{^2}\Pi_\Omega$ in \ce{C6H}.

\subsection*{Band XVII: B${^2}\Pi_\Omega - $X${^2}\Pi_\Omega$ vibronic band}
\begin{figure}[ht!]
  \centering
  \includegraphics[width=0.48\textwidth]{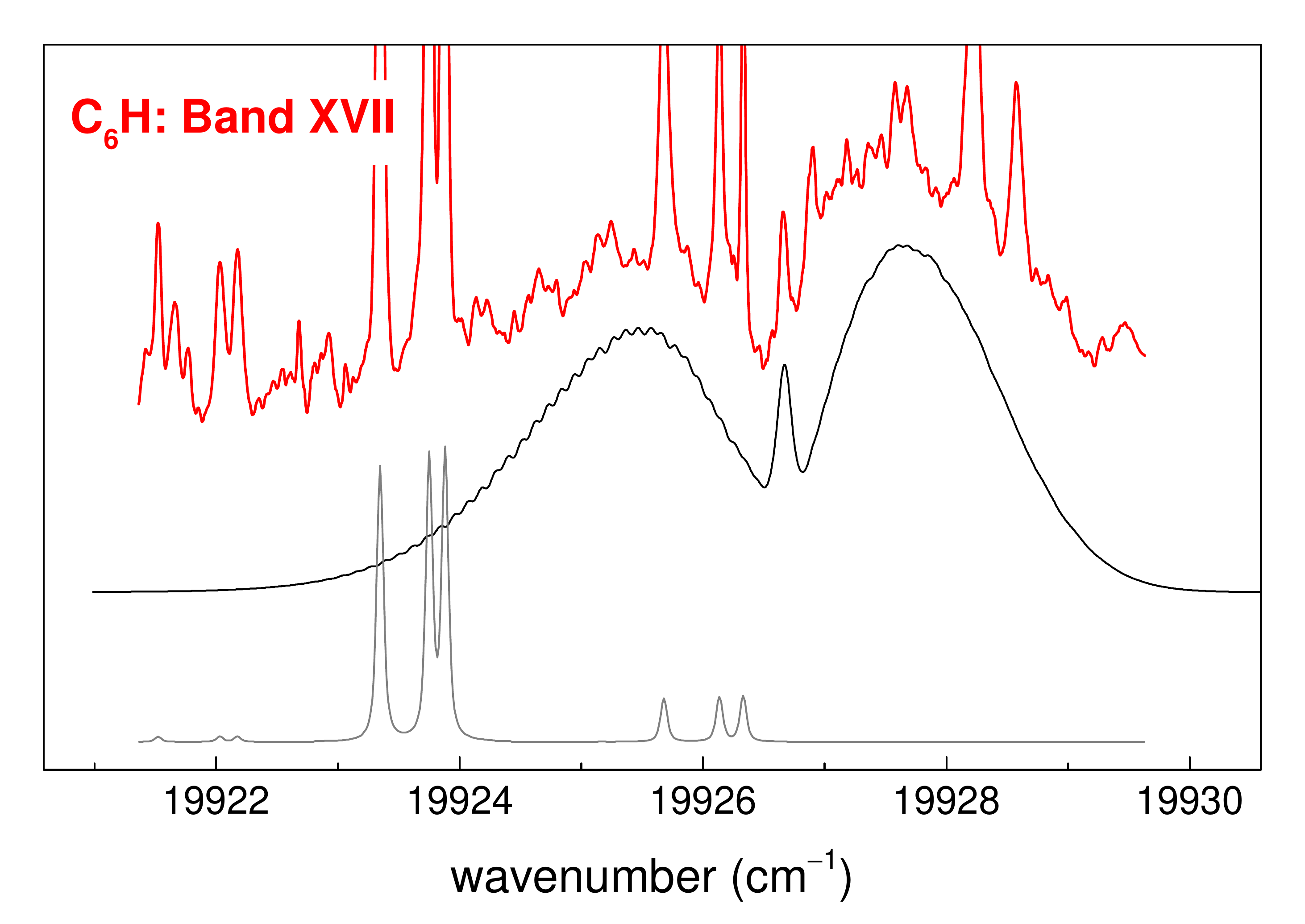}
  \caption{Band XVII of \ce{C6H} measured with in the slit jet plasma configuration and a simulation based on a $^2\Pi_\Omega - {^2}\Pi_\Omega$ band type. The narrow absorption peaks arise from \ce{C2} $(0,0)$ and $(1,1)$ Swan bands.}
  \label{fig:Band-XVII}
\end{figure}

The band system in the frequency range $19\,922 - 19\,930$ \wn, displayed in Fig.~\ref{fig:Band-XVII}, is of low quality in terms of signal-to-noise and coverage with sharp resonances associated with C$_2$. The assignment of this feature is very tentative: the central and rather narrow peak at $19\,926.7$ \wn\ might be considered as a $Q$~branch and indicative of a $^2\Pi_{3/2}-{^2}\Pi_{3/2}$ component. But using a similar argument as with Band VII, based on the position of the $Q$~branch, the spectrum can be simulated by superimposing the two spin-orbit components. Only rough estimates of the $T_0$, $B'$, and $A'$ constants can be given on the basis of a contour analysis, and the corresponding values are listed in Table~\ref{tab:summary}.

\subsection*{Bands XVIII and XIX}
\begin{figure}[ht!]
  \centering
  \includegraphics[width=0.48\textwidth]{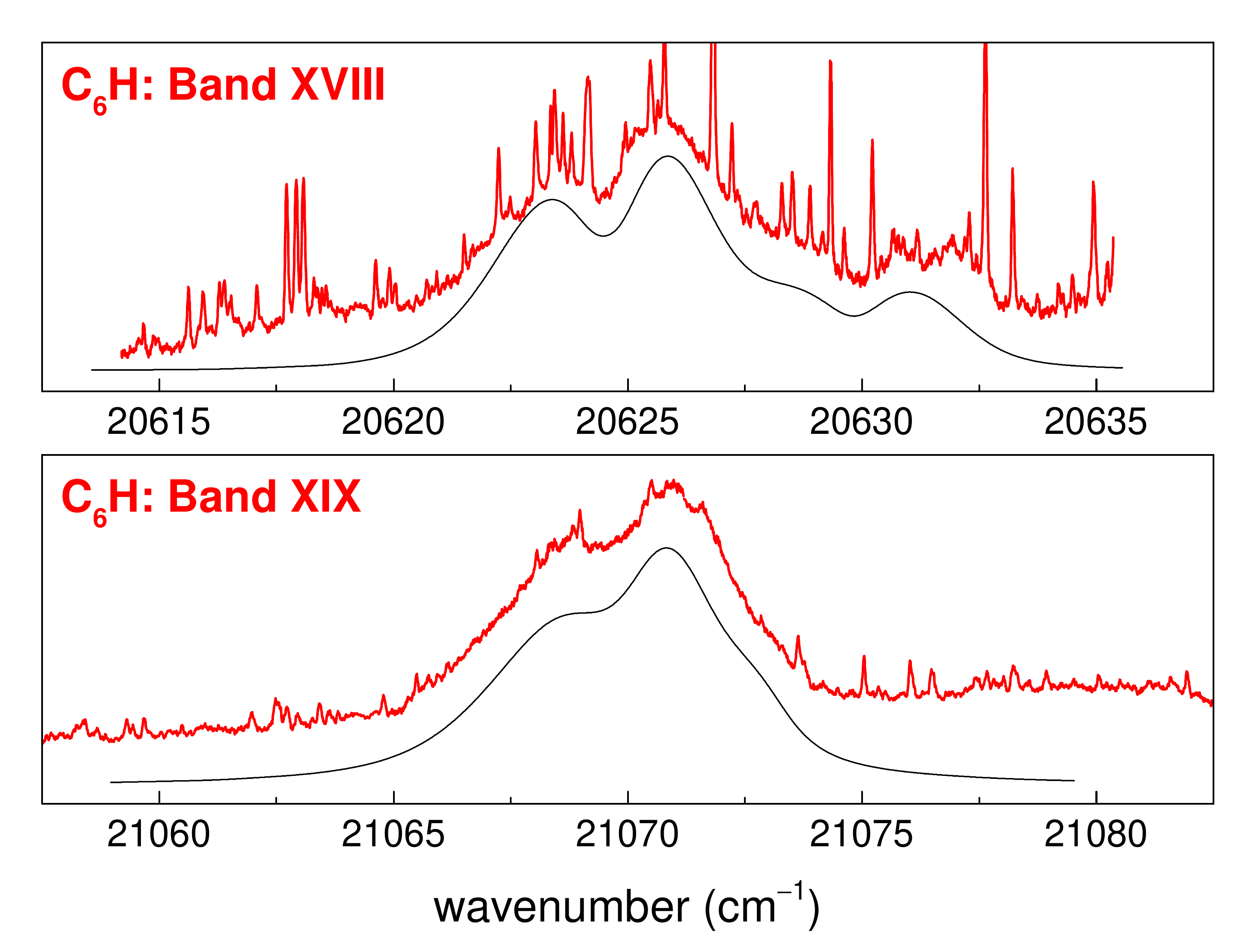}
  \caption{Bands XVIII and XIX of \ce{C6H} measured with a slit plasma nozzle and a contour simulation of the spectrum (in black). Note that the wavenumber axis is wider than in the previous figures.}
  \label{fig:Band-v5-v3}
\end{figure}

Finally, in the highest energy region investigated in the present study, in the frequency range above $20\,000$~\wn, two bands have been observed, which are shown in Fig.~\ref{fig:Band-v5-v3}. Despite the two bands having been measured with a slit plasma nozzle, individual rotational lines of \ce{C6H} could not be resolved. The broadening is effective to the extent that the $^2\Pi_{3/2}$ and $^2\Pi_{1/2}$ components are somewhat overlapped, without a distinct $Q$~branch visible. Moreover, the spectra, in particular for Band XVIII, are overlaid by a congested forest of narrow resonances. Assuming a rotational temperature of $\sim$~22~K, and fixing the ground state constants to those of the $^2\Pi$ ground state of \ce{C6H}, a band contour simulation can be produced yielding the molecular constants listed in Table~\ref{tab:summary}.

\section*{Discussion}
In the present study 19 vibronic (sub)bands pertaining to the B$^2\Pi -$X$^2\Pi$ electronic system of the \ce{C6H} molecule have been observed. For the \ce{C6D} isotopologue the seven lowest energy bands have been observed and for \ce{^{13}C6H} the eight lowest bands. From the fitting of the recorded spectra, values are derived for the band origins $T_0$, effective rotational constants $B_\mathrm{eff}$, effective spin-rotation constants $\gamma_\mathrm{eff}$, and, for the case of ${^2}\Pi$ vibronic states, a spin-orbit coupling constant $A_\mathrm{SO}$. For some of the bands, clearly rotationally resolved spectra have been measured and in these cases a fit has been made comparing individual line positions (see \href{dx.doi.org/10.1021/acs.jpca.6b06647}{Supporting Information} for the line lists). In some additional cases spectra are unresolved or strongly overlaid with narrow spectral features belonging to CH and C$_2$ radical species. In such cases only a band contour simulation has been produced, for which the uncertainty on the derived spectroscopic parameters is larger.

All the bands investigated are connected to an electronic transition of $^2\Pi - {^2}\Pi$ character. If the \ce{C6H} molecule were to remain linear in ground and excited state levels then the vibronic absorption bands could all be described as $^2\Pi - {^2}\Pi$ transitions in a diatomic molecule. Even if both $^2\Pi$ states would fit a Hund's case(b) description this would remain the case.~\cite{Herzberg1966} If however the spin-orbit splitting in a $^2\Pi$ state becomes appreciable, then the bending vibrations couple to the electronic motion, giving rise to doubly degenerate structures and bands that appear as $^2\Sigma - {^2}\Sigma$ transitions --- a phenomenon known as the Renner-Teller (RT) effect. This holds for the case of a single bending vibration. For the non-degenerate stretching vibrations, this is not the case and the electronic absorption bands retain the character of a $^2\Pi - {^2}\Pi$ transition. The same holds for a combination of two degenerate bending vibrations that will also display the character of a $^2\Pi - {^2}\Pi$ transition.

The \ce{C6H} molecule is the first member of the C$_n$H series to be linear in the X$^2\Pi$ ground state, in contrast to the smaller chains like C$_4$H exhibiting a $^2\Sigma^+$ ground state. This was shown in electronic structure calculations~\cite{Pauzat1989,Sobolewski1995} and found to be consistent with radio astronomical observations.~\cite{Cernicharo1987b} \ce{C6H} is also linear in the vibrationless B$^2\Pi$ excited state as is derived from the gas-phase observation of the B$^2\Pi-$X$^2\Pi$ origin band spectrum.~\cite{Linnartz1999} It has 16 vibrational modes, of which there are 5 doubly degenerate bending modes (in the two perpendicular planes containing the molecular axis) and 6 non-degenerate stretching modes. Four different \emph{ab initio} studies have been reported that calculated the non-degenerate stretching modes along the linear axis of the C$_6$H backbone.~\cite{Doyle1991,Liu1992,Brown1999,Cao2001} The calculated eigen frequencies of the fundamental modes, as listed in Table~\ref{tab:Eigenfrequencies}, are in reasonable agreement with each other. Two of the cited works on \emph{ab initio} calculations~\cite{Brown1999,Cao2001} also include results on the five doubly degenerate bending modes. Energies of the degenerate modes were calculated to exhibit mode frequencies of $100-600$~\wn, while the energies of the non-degenerate modes fall in the range $600-3\,500$~\wn. 

\begin{table}[H]
\centering
\caption{The calculated eigen frequencies of the six non-degenerate stretching modes of \ce{C6H} as in the four reported \emph{ab initio} studies~\cite{Doyle1991,Liu1992,Brown1999,Cao2001} together with the vibrational frequencies derived from the observed gas-phase vibronic spectrum. The observed frequency at $1\,953.4$~\wn\ (marked by an $*$) is taken from Ref. [\!\!\citenum{Doyle1991}\,]. All values in~\wn.}
\label{tab:Eigenfrequencies}
\resizebox{\columnwidth}{!}{
\begin{tabular}{@{}cccccc@{}}
\toprule
Mode            & Doyle 1991~\cite{Doyle1991} & Liu 1992~\cite{Liu1992} & Brown 1999~\cite{Brown1999} & Cao 2001~\cite{Cao2001} & Observed \\ \midrule
$\nu_1$      & 3\,659.1     & 3\,609.0   & 3\,448       & 3\,457     &          \\
$\nu_2$      & 2\,436.7     & 2\,197.8   & 2\,109       & 2\,137     & 1\,953.4*  \\
$\nu_3$      & 2\,336.0     & 2\,139.6   & 2\,080       & 2\,105     & 2\,080.8   \\
$\nu_4$      & 2\,092.0     & 1\,974.5   & 1\,872       & 1\,895     & 1\,637.4   \\
$\nu_5$      & 1\,213.0     & 1\,201.8   & 1\,210       & 1\,224     &    \\
$\nu_6$      &  653.9     &  638.5   &  642       &  650     & 658.9    \\ \bottomrule
\end{tabular}
}
\renewcommand{\mpfootnoterule}{}
\end{table}

\begin{figure}[]
  \centering
  \includegraphics[width=\columnwidth]{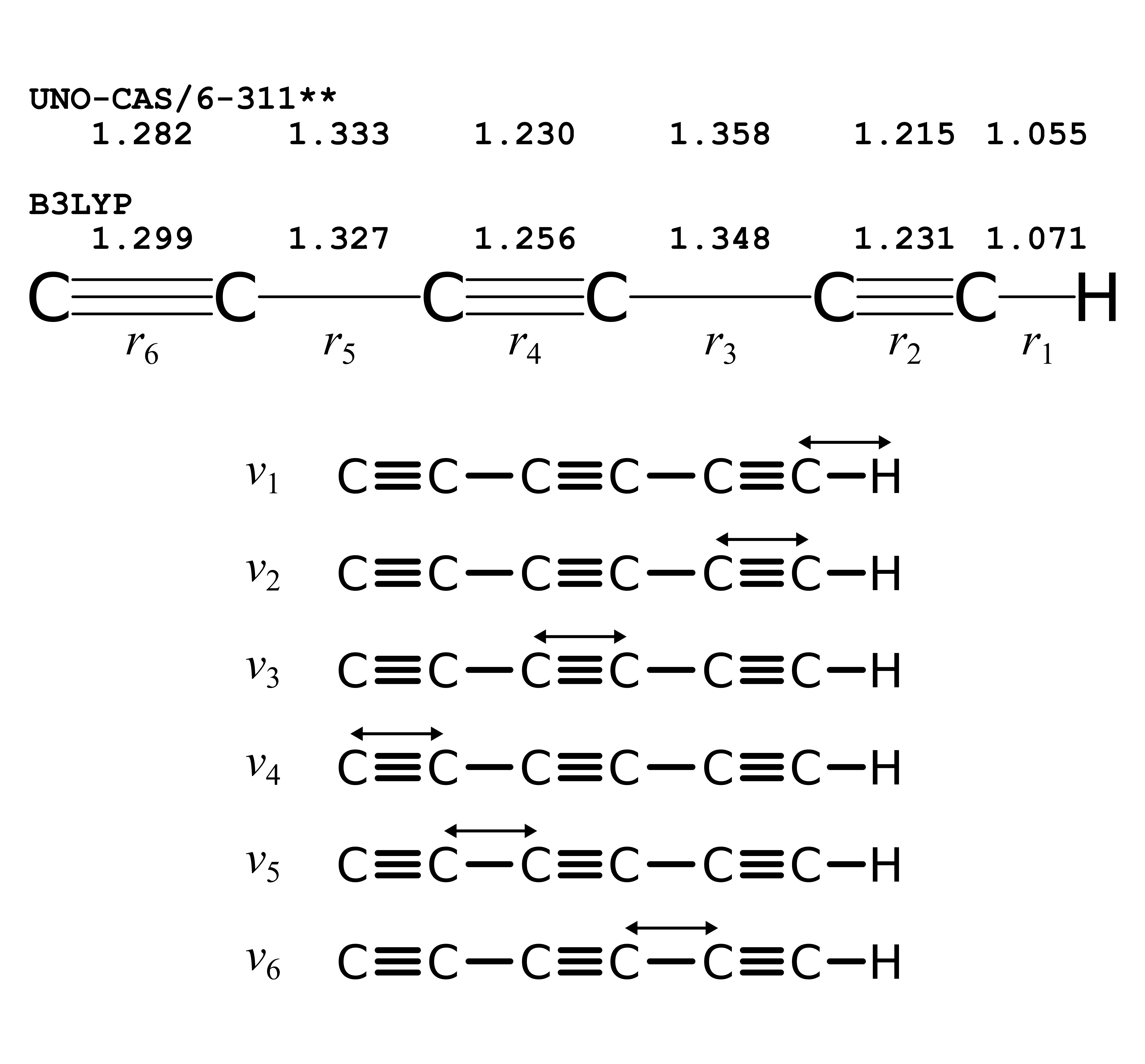}
  \caption{The geometry of the \ce{C6H} molecule, adopted from the \texttt{UNO-CAS/6-311**} calculations in Ref. [\!\!\citenum{Liu1992}\,] and from the more recent \texttt{B3LYP} calculations in Ref. [\!\!\citenum{Brown1999}\,]. Bond lengths (in~\AA) are exaggerated to emphasize their differences. Also shown are the six fundamental non-degenerate eigen modes where double arrows indicate on which bond the stretching vibration primarily occurs.}
  \label{fig:C6Hbonds}
\end{figure}

In Fig.~\ref{fig:C6Hbonds} the fundamental non-degenerate eigen modes are drawn in the geometry of the C$_6$H molecule, including calculated bond lengths.~\cite{Liu1992,Brown1999} The highest frequency mode, $\nu_1$, has the typical frequency of a C--H stretching mode and is assigned as such. The three subsequent modes, $\nu_2$, $\nu_3$, and $\nu_4$, in order of decreasing frequency, are associated with the C$\equiv$C triple bonds (Fig.~\ref{fig:C6Hbonds}) where the highest frequency is assigned to the mode with the shortest bond length among the three. The two lowest frequencies, $\nu_5$ and $\nu_6$, are associated with the vibrational motion centered on the C--C single bonds.

These calculations are taken as the basis for assigning the recorded vibronic bands. It should be noted that all four \emph{ab initio} calculations were performed for the ground state electronic configuration, and that we assume the vibrational level structure of the B$^2\Pi$ excited state to reproduce that of the ground state. This is, of course, a rather strong but common assumption. It is noted further that in the study of Brown et al.,~\cite{Brown1999} ambiguity arises on the identification of the infrared mode observed in the ground state of C$_6$H at $1\,953.4$~\wn,~\cite{Doyle1991} which is assigned as $\nu_3$ in one of their tables. In another table in the same paper, they still assign this mode to $\nu_3$ but they list a calculated frequency of $2\,109$~\wn, which corresponds to $\nu_2$ in their previous table. In any case, this IR-absorption is identified as the C$\equiv$C stretching vibration next to the C--H bond ($r_2$ in Fig.~\ref{fig:C6Hbonds}). In addition, in the photodetachment study by Taylor et al.~\cite{Taylor1998} a vibrational mode at $2\,202$~\wn\ was deduced, which however is associated with a low-lying electronic state of $^2\Sigma^+$ symmetry.~\cite{Sobolewski1995} While the two bands mentioned pertain to the X$^2\Pi$ and $A^2\Sigma^+$ states, in the matrix studies two bands were observed at 21\,645~\wn\ and 22\,517~\wn, corresponding to vibrational excitations of 2\,791~\wn\ and 3\,664~\wn\ in the B$^2\Pi$ state. These bands were tentatively assigned as $\nu_3 + \nu_6$ and $\nu_3 + \nu_5$ combination bands,~\cite{Freivogel1995} while based on the consistent \emph{ab initio} calculations~\cite{Liu1992,Brown1999,Cao2001,Doyle1991} the higher frequency band might also be associated with $\nu_1$.

All observed bands originate from the ground vibrational level, starting from the lowest lying X$^2\Pi_{3/2}$ level, or the X$^2\Pi_{1/2}$ spin-orbit component which is excited by $\sim$~15.0~cm$^{-1}$, or from the $\mu^2\Sigma$ component of the RT manifold, which is at an excitation energy of $10.7 \pm 0.5$ cm$^{-1}$ above the ground level (X$^2\Pi_{3/2}$).~\cite{Zhao2011b} The $\mu^2\Sigma$ component is associated with the lowest $\nu_{11}=1$ bending vibration in the electronic ground state, which vibronically couples to produce a pair of $^2\Sigma$ states, referred to as a lower $\mu^2\Sigma$ component and a higher lying $\kappa^2\Sigma$ component. This $\kappa^2\Sigma$ component has not yet been observed spectroscopically. Also a pair composed of the $^2\Delta_{5/2}$ and $^2\Delta_{3/2}$ levels follows from the RT coupling, excited at $\sim$~50~cm$^{-1}$, which has been observed for \ce{C6H} in a microwave study.~\cite{Gottlieb2010} In the present study no signature of this $^2\Delta$ pair can be identified.

From a spectroscopic analysis of the rovibronic bands and their treatment in the previous section it appears that all bands can be described either by  a $^2\Pi - {^2}\Pi$ or a $^2\Sigma - {^2}\Sigma$ type transition. In the following, all available data are merged with the aim to assign the observed bands to the vibronic level structure of the B$^2\Pi$ state.

\subsection*{The $^2\Pi$ levels}
In all cases the bands connected to $^2\Pi - {^2}\Pi$ character display the characteristic spin-orbit splitting  with $A'' \approx -15$~\wn\ for the \ce{C6H} and \ce{C6D} isotopologues, hence yielding an inverted spin-orbit level ordering between $^2\Pi_{3/2}$ and $^2\Pi_{1/2}$. As discussed previously,~\cite{Bacalla2015} the ground state spin-orbit splitting for \ce{^{13}C6H} is $A'' \approx -12$~\wn.
In all the fits and band contour simulations the ground state spin-orbit constants have been kept fixed at these values. The spin-orbit structure of the B$^2\Pi -$X$^2\Pi$ (0,0) origin bands of \ce{C6H} and \ce{C6D} was discussed in earlier studies~\cite{Linnartz1999} and yielded an excited state spin-orbit constant of $A' \approx -24$~\wn. The value for \ce{^{13}C6H} of $A' \approx -21$~\wn\ can be considered consistent.~\cite{Bacalla2015}

The spectra of the $^2\Pi - {^2}\Pi$ bands are compared with observations from matrix isolation data, which bear the advantage that a mass-analyzer stage selects the ionized radical species before depositing in a matrix. In this way the matrix data provide an additional constraint that can be used to identify the carriers of the bands discussed here. For C$_6$H such studies were typically performed in neon matrices cooled to 5 K, where only the $^2\Pi$ ground state was populated.~\cite{Forney1995,Freivogel1995} In these matrix studies, no excitations associated with the Renner-Teller $\nu_{11}$ bending mode were observed, thus creating another helpful selection criterion. Furthermore, several band features, observed in the matrix spectra, were assigned to vibrational modes in the electronically excited C$_6$H molecule, from a comparison with calculated fundamental frequencies for the electronic ground state of the molecule as available at the time.~\cite{Liu1992,Brown1999} Besides the $0_0^0$ origin band of the B$^2\Pi -$X$^2\Pi$ system, three additional non-degenerate modes were observed at vibrational excitation energies of 2\,080.8~\wn, 1\,637.4~\wn, and 658.9~\wn. These stretching vibrations generate a $^2\Pi - {^2}\Pi$ type transition as is clearly observed in the spectra. Of importance for the present analysis is that the matrix shift for the $0_0^0$ origin band amounts to $-136$~\wn. A typical finding from matrix isolation spectroscopy is that matrix shifts tend to be constants for vibronic structures within a few \wn. This forms another ingredient for identifying bands. 

The spectra of Bands XV and XVI, for which both $^2\Pi_{3/2} - {^2}\Pi_{3/2}$ and $^2\Pi_{1/2} - {^2}\Pi_{1/2}$ components were observed for \ce{C6H}, can be unambiguously assigned to a  $^2\Pi - {^2}\Pi$ doublet for a number of reasons, even though the spectrum is heavily contaminated. This band was observed in the matrix study~\cite{Freivogel1995} exhibiting a matrix shift of $-137$~\wn\ with respect to the present measurement as well as to the shift observed in a lower resolution gas-phase discharge spectrum.~\cite{Kotterer1997} The present study yields an excited state spin-orbit constant of $A'=-24.4$~\wn, which is consistent with that of the origin band. We follow the assignment as in previous studies to match the two components of this band to the lowest non-degenerate vibration, the $\nu_6$ mode, which derives its energy from the vibrational motion of a \ce{C-C} single bond in the molecule (cf. Fig. \ref{fig:C6Hbonds}). It is noted, however, that the intensity ratio between the $^2\Pi_{3/2} - {^2}\Pi_{3/2}$ and the $^2\Pi_{1/2} - {^2}\Pi_{1/2}$ spin-orbit components behaves somewhat irregularly (see the Results section for Bands XV and XVI).

In a similar way, the bands observed for \ce{C6H} in the higher energy region (Bands  XVIII and XIX)  can be well reproduced by a $^2\Pi - {^2}\Pi$ doublet with excited state spin-orbit values of $A'=-20.3$~\wn\ and $A'=-16.8$~\wn, which are in good agreement with the $A'$ spin-orbit constant for the vibrationless origin band. Comparison of the present gas-phase band origins with matrix data yields matrix shifts  $\sim$~$133$ and $\sim$~$142$~cm$^{-1}$, in good agreement with the typical matrix shift as measured for the $0^0_0$ origin band. In previous matrix isolation studies~\cite{Forney1995,Freivogel1995} these bands were assigned to $\nu_5$ and $\nu_3$ modes from a comparison with the then available \emph{ab initio} data.~\cite{Liu1992} In the later calculations~\cite{Brown1999,Cao2001} the $\nu_3$ mode is shifted to lower frequencies, which is indeed in agreement with the present observation of $\Delta\nu=2\,080.8$~\wn. For the assignment of the band observed at $\Delta\nu=1\,637.4$~\wn, the situation is less clear. In view of the most recent calculations this band matches better with the $\nu_4$ mode, than with the earlier identification as $\nu_5$.~\cite{Freivogel1995} Hence we keep this assignment as tentative at $\nu_4$.

The observed spectra (see Fig.~\ref{fig:Band-v5-v3}), recorded for the first time in the gas phase, are markedly different from all other bands measured in the sense that in all other cases resolved rotational lines appear. The $\nu_3$ and $\nu_4$ excited state vibrational modes exhibit a triple-bond stretching vibration (cf. Fig.~\ref{fig:C6Hbonds}), which means that the molecule does not deviate from linearity. Hence, a rotational line separation is expected to be very similar to that of other $^2\Pi - {^2}\Pi$ bands and rotational lines should be resolved in the same manner as for the other bands detected. Because this is not observed, it is assumed that the $\nu_3$ and the $\nu_4$ modes undergo lifetime broadening, most likely due to internal conversion. It is noted that the next carbon chain molecule in the C$_{2n}$H series, \ce{C8H}, also exhibits lifetime broadening already in its origin band.~\cite{Birza2003} For the moment, we leave the latter issues as outside the scope of the present work. Fluorescence studies, similarly as performed for the C$_4$H molecule,~\cite{Hoshina1998} might clarify some of the dynamics.

Since the stretching vibrations (which give rise to a state of total $^2\Pi$ symmetry) are all predicted upward from $\sim$~$640$~\wn, this leaves little space for assigning the other low frequency modes of total $^2\Pi$ symmetry. What remains to be considered are the following bands: Bands VII through XI and Bands XIII and XIV, while Band XVII is a special case, since its frequency is larger than the 650~\wn\ demarcation point between stretching and bending modes in C$_6$H.

Bands IX and X, detected for \ce{C6H} only, reproduce the characteristic structure of a $^2\Pi_{\Omega} -$X$^2\Pi_{\Omega}$ doublet. The derived value of $A'=-23.77$~\wn\ corresponds to the typical value for the spin-orbit constant also found for the $0^0_0$ origin band and the $\nu_6$ mode (Bands XV and XVI). In the neon matrix spectra~\cite{Freivogel1995} this band was observed at $19\,286$~\wn. The presently observed gas-phase B$^2\Pi_{\Omega} -$X$^2\Pi_{\Omega}$ doublet is thus shifted by $143$~\wn, consistent with the typical gas-matrix shift for \ce{C6H}. This mode has an excitation energy of $439.5$~\wn.

Bands XIII and XIV again display the typical characteristics of a $^2\Pi-{^2}\Pi$ doublet, but this assignment yields an excited state spin-orbit constant of $A'=-8.0$~\wn, which is atypical for the B$^2\Pi$ state. The vibrational mode has an excitation energy of 603~\wn\ above the band origin.

Band VIII, observed for both \ce{C6H} and \ce{^{13}C6H}, and Band XI observed only for \ce{C6H}, both display a characteristic of a  $^2\Pi_{3/2}-{^2}\Pi_{3/2}$ component. In view of the fact that rotational fits could be made for the \ce{C6H} spectra, resulting in a rotational constant  matching that of \ce{C6H}, the assignment of Bands VIII and XI to the \ce{C6H} molecule can be considered as unambiguous. Band VIII has, in \ce{C6H}, an excitation energy of 337~\wn\ above the origin, while in \ce{^{13}C6H} it is at 324~\wn, consistent with the pattern of isotope shifts for the vibrational bands.

Band VII, observed for all three isotopologues, involves a $^2\Pi_{3/2}-{^2}\Pi_{3/2}$ component. The hypothesis that this component is overlaid with a $^2\Pi_{1/2}-{^2}\Pi_{1/2}$ component yields an upper state spin-orbit constant of $A'=-14.6$~\wn. This interpretation should however be considered as tentative. In any case, the assignment of the $^2\Pi_{3/2}-{^2}\Pi_{3/2}$ component may be considered as firm, as is the assignment of these bands to the \ce{C6H} molecule. Band VII lies at an excitation energy of $309$~\wn\ for \ce{C6H}, and at $296$~\wn\ for \ce{^{13}C6H}, again consistent with the isotopic pattern.

As for the vibrational levels of these bands, the modes with frequencies 308.8~\wn, 336.7~\wn, 439.5~\wn, and 603.3~\wn, hence below $\sim$~650~\wn, cannot be assigned to non-degenerate eigen modes. There simply are no stretching modes of such low energies predicted in the \emph{ab initio} calculations~\cite{Doyle1991,Liu1992,Brown1999,Cao2001} (see Table \ref{tab:Eigenfrequencies}). Hence, the entire set of low-lying $^2\Pi$ states must be due to combination modes of bending vibrations, such as $2\nu_{11}$, $\nu_{11}+\nu_{10}$, $2\nu_{10}$, etc.

The observed spectrum of Band XVII is weak and only a band contour simulation has been possible. Its excitation energy of 937~\wn\ does not match with a stretching eigen mode in the \emph{ab initio} calculations. The $\nu_5$ stretching mode yields a consistent value close to $\sim$~$1\,210$~\wn\ in the \emph{ab initio} calculations (cf. Table \ref{tab:Eigenfrequencies}), which is off by $\sim$~300~\wn\ with respect to the band origin of Band XVII. If $\nu_5$ is not covered by the band observed at 1\,637.4~\wn\ (see above) then it might tentatively be assigned to the excitation at 937~\wn. Improved calculations of the B$^2\Pi$ excited electronic state vibrations  of \ce{C6H} might decide on such a tentative assignment. Alternatively, and more probably, this feature could be due to a combination band, e.g., to a $2\nu_9$ mode. The observation that the $B'$ rotational constants of several of these vibronic modes are larger than that of the $0^0_0$ origin band implies that the vibration-rotation interaction is negative. This is typical for bending or bending-like modes, as was shown in the example of vibrations in the ground state of HC$_6$H.~\cite{Chang2016} This lends further credit to assigning the low-lying $^2\Pi$ states to bending combination modes. 

\subsection*{The $^2\Sigma$ levels}
In order to assign the remaining features in the recorded spectra interpreted as $^2\Sigma-{^2}\Sigma$ type bands, the interaction between degenerate electronic states and its fundamental bending modes, i.e., the Renner-Teller effect, has to be analyzed for a number of vibronically excited states of $^2\Sigma$ symmetry. In two sets of \emph{ab initio} calculations performed for the electronic ground state,~\cite{Brown1999,Cao2001} some 5 doubly degenerate bending modes were calculated (using \texttt{B3LYP} hybrid functionals in Ref. [\!\!\citenum{Brown1999}\,]): $\nu_{11}$ at 110-120 \wn, $\nu_{10}$ at 214--255~\wn, $\nu_9$ at 396--445~\wn, $\nu_8$ at 519--553~\wn, and $\nu_7$ at 561--679~\wn. Each of these give rise to a $\mu^2\Sigma - \kappa^2\Sigma$ doublet on the basis of which the five observed $^2\Sigma - {^2}\Sigma$ transitions are to be assigned.

The Hamiltonian describing the energy of vibronic states is given as the sum of the energies involved in pure vibration, in the RT coupling, and in the spin-orbit (SO) interaction: $$H = H_\mathrm{vib} + H_\mathrm{RT} + H_\mathrm{SO}.$$ With the measured band origin $T_0$ for a particular transition, diagonalization of a matrix for calculated matrix elements for the various contributions will yield values of the true spin-orbit coupling constant ($A_\mathrm{SO}'' = -47.5$~\wn\ and $A_\mathrm{SO}' = -23.74$~\wn) as well as for the Renner-Teller parameter $\epsilon$, derived typically in the product $\epsilon\omega$,~\cite{Zhao2011b} with $\omega$ as the vibrational frequency. Then for ${^2}\Sigma$ vibronic states, the effective splitting $A_\Sigma$ between the two ${^2}\Sigma$ components (labeled as $\mu{^2}\Sigma$ and $\kappa{^2}\Sigma$) is given by:

\begin{equation}
\label{eq:SO-splitting}
A_\Sigma = \sqrt{{A_\textrm{SO}}^2 + 4\epsilon^2\omega^2}
\end{equation}

\noindent and the effective rotational constants can be approximated as:~\cite{Hougen1962,Herzberg1966}

\begin{equation}
B_{\textrm{eff}}^{\mu/\kappa} = B \left( 1 \pm \frac{{A_{\textrm{SO}}}^2 B}{{A_{\Sigma}}^3} \right)
\end{equation}
\\
\begin{equation}
\gamma_{\textrm{eff}}^{\mu/\kappa} = 2B \left( 1 - \frac{2\epsilon\omega}{A_{\Sigma}} \pm \frac{{A_\textrm{SO}}^2 B}{{A_{\Sigma}}^3} \right).
\end{equation}
\\
Such a RT analysis can be performed for the two low-lying bending modes $\nu_{10}$ and $\nu_{11}$. For the lowest $\nu_{11}=1$ vibrational mode, this was previously done for the \ce{C6H} and \ce{C6D} molecule, forming the basis for an assignment of a $\{11\}^1_1\ {\mu}^2{\Sigma}-{\mu}^2{\Sigma}$ transition.~\cite{Zhao2011b} A large RT parameter was obtained for the ground state ($\epsilon_{11}'' \sim$~0.99), while a much smaller value was found for the excited state ($\epsilon_{11}' \sim$~0.09).

The presently observed Band III of \ce{^{13}C6H} can be assigned in an analogous manner. On the basis of isotopic regularity, no other band feature is observed between this band and the (0,0) origin band of $^2\Pi - {^2}\Pi$ signature (combined Bands I and II). As listed in Table \ref{tab:summary}, the isotopic shifts of the first three observed bands of \ce{^{13}C6H} are decreasing; Band III is consistent with this trend, having a band shift of 0.1~cm$^{-1}$ to higher energy compared to the regular isotopologue \ce{C6H}. In addition, its band contour appears very similar as in the other isotopologues. Together with the results of the rotational line fit, all of these support for an unambiguous assignment of Band III in $^{13}$C$_6$H.

Furthermore, the RT analysis for the $\nu_{11}=1$ mode in the upper electronic state predicts energy levels in the form of $^2\Sigma$ and $^2\Delta$ pairs, with the $\{11\}^1\ {\kappa}^2{\Sigma}$ level lying $\sim$~30~cm$^{-1}$ above the $\{11\}^1\ {\mu}^2{\Sigma}$ level. No bands with nearly similar features have been observed in the predicted energy range, for neither of the three isotopologues. In general, a stronger RT effect in the upper state weakens the $\kappa^2{\Sigma} - {\mu}^2{\Sigma}$ transition which indicates a lesser coupling intensity between vibrational and electronic orbital angular momenta. In case of a strong Renner-Teller effect, the $\Pi$ electronic orbital is coupled to a vibrational angular momentum (of $\Pi$ symmetry), resulting into two $\Sigma$ and one $\Delta$ vibronic states, where the two $\Sigma$ states have different parity signs ($^2\Sigma^+$ and $^2\Sigma^-$). The transition to a pure $^2\Sigma^-$ state, from the $\mu^2\Sigma^+$ ground state becomes forbidden.~\cite{Hougen1962,Herzberg1966} Apparently, the value of $\epsilon_{11}'$ is large enough to prohibit an observable intensity in the $\kappa^2\Sigma$ component, and indeed no band with a substantial S/N is found at the expected position for neither of the three isotopologues. Transitions to the $^2\Delta_{5/2}$ and $^2\Delta_{3/2}$ also cannot be observed since the only populated vibronic component of the ground $\nu_{11}=1$ vibrational mode is the $\{11\}_1\ {\mu}^2{\Sigma}$ state and the selection rule does not allow for a change in vibronic symmetry greater than by one quantum number.

For the low-lying $\nu_{10}=1$ bending mode in the B$^2\Pi$ excited state, a similar RT analysis can be performed yielding a $\mu$-$\kappa$ doublet of $^2{\Sigma}$ symmetry. Experimentally, from the characteristic shapes of Bands IV and V, one can infer that they start from the same ground state, and that they are excited to coupled vibronic levels. Indeed, from the rotational fit, both of these bands can be simulated using the same ground state constants which are those of the $\{11\}_1\ {\mu}^2\Sigma$ state. Moreover, the $B'$ rotational constants for both bands are rather similar, as expected for a RT doublet.

The observed difference between the two band origins ($\sim$~23~cm$^{-1}$) corresponds to the splitting $A_\Sigma^\mathrm{obs}$ between the $\mu^2\Sigma$ and the $\kappa^2\Sigma$ doublet. In contrast to the case of the $\nu_{11}$ bending mode, the combination of Bands IV and V may be interpreted as a doublet of transitions to the $\{10\}^1\ {\mu}^2{\Sigma}$ lower and $\{10\}^1\ {\kappa}^2{\Sigma}$ upper states of the RT doublet. If the value of the true spin-orbit constant $A_\mathrm{SO}'=-23.74$~\wn, as determined from the analysis of $\nu_{11}$ is adopted, then an effective splitting  between the band origins is calculated at $|A_\Sigma^\mathrm{calc}| > |A_\Sigma^\mathrm{obs}|$, via Eq.~(\ref{eq:SO-splitting}). This hints at the one side to a very small Renner-Teller parameter ($\epsilon_{10}' < 0.01$) but still, the splitting between band origins for $\{10\}^1\ \kappa^2\Sigma$ and $\{10\}^1\ \mu^2\Sigma$ is too small for a consistent analysis. It is likely that the $\{10\}^1\ \kappa^2\Sigma$ component (Band V) is shifted downwards, for example, due to a perturbation or a Fermi resonance. The fact that the $\gamma'$ parameter for Band V, specifically for the \ce{C6H} isotopologue ($\gamma' = 0.107$~\wn), corresponds to $\gamma' > 2B'$ makes it unphysical and also points at a perturbation. In any case, the $\epsilon_{10}'$ value is expected to be small because both RT components are observed in the spectrum for all three isotopologues.

The next band observed, Band VI, also exhibits the character of a $^2\Sigma - {^2}\Sigma$ transition. Here, as in the case of the lowest $\nu_{11}$, no second component in a $\mu$-$\kappa$ RT pair is observed. In view of the numbering of bending vibrations, Band VI may be assigned as the lower $\mu$ component of $\nu_9$. In principle, the next band of $^2\Sigma - {^2}\Sigma$ character (Band XII) might be considered as the upper $\kappa$ component of $\nu_9$. However, the $B'$ rotational constants for these bands differ substantially while they should be similar in a RT pair. For this reason, the last $^2\Sigma - {^2}\Sigma$ transition (Band XII) is tentatively assigned as the $\mu$ component of $\nu_8$.

\subsection*{Isotope shifts}
Table~\ref{tab:summary} displays values for the isotope shifts for the bands observed for \ce{^{13}C6H} and \ce{C6D} with respect to those of \ce{C6H}. It should be noted that isotope shifts are determined only for degenerate bending modes, that is, for the $^2\Sigma - {^2}\Sigma$ vibronic bands or for combinations thereof. The values can be understood by assuming that the effect of zero-point vibration makes up for a decisive contribution. If the ground state potential hypersurface is considered to be deeper than for the excited state, i.e., stronger chemical binding in the ground state, this explains the blue shift of the origin bands for the heavier molecules. The effect appears to be much stronger for \ce{C6D}  than for \ce{^{13}C6H} in view of the larger mass difference between D and H. For gradually higher vibrational excitation a regular effect of redshift for higher modes is observed of overall $\sim$~10~\wn\ for eigen modes up to $\sim$~300~\wn. While for \ce{^{13}C6H} the shift pattern develops smoothly and linearly, for \ce{C6D} the pattern is less regular. There is an outlier for Band V, which may be associated with the perturbation found in the Renner-Teller splitting of Bands IV and V. A perturbation-induced down shifting of Band V in \ce{C6H} will result in an increased isotope shift for \ce{C6D} and a decreased isotope shift for \ce{^{13}C6H} (cf. Fig.~\ref{fig:compiled-allbands}). Indeed, the isotope shift of Band V for  \ce{^{13}C6H} deviates somewhat ($\sim 2$ \wn) from a smooth regular progression of isotope shifts.

The measured isotope shifts in the present study all pertain, except for the origin band, to degenerate vibrational modes or combinations thereof. It is found that isotopic shifts for these bands in \ce{^{13}C6H} and \ce{C6D} behave very differently, while in \ce{C6D} the isotope shift is $\sim$~50~\wn, for \ce{^{13}C6H} this is typically $\sim$~$-5$~\wn, hence of opposite sign. Brown et al.~\cite{Brown1999} have performed specific isotope shift calculations for the non-degenerate $\nu_2$ vibrational mode (reassigned from $\nu_3$ in their paper) yielding a similar value of 78~\wn\ for both isotope shifts of \ce{^{13}C6H} and \ce{C6D} (with respect to \ce{C6H}). This finding seems difficult to reconcile with the present measurements. Refined isotope shift calculations for a variety of bands may settle this issue.

\balance

\section*{Conclusion}
\begin{figure*}[h]
  \centering
  \includegraphics[width=0.9\textwidth]{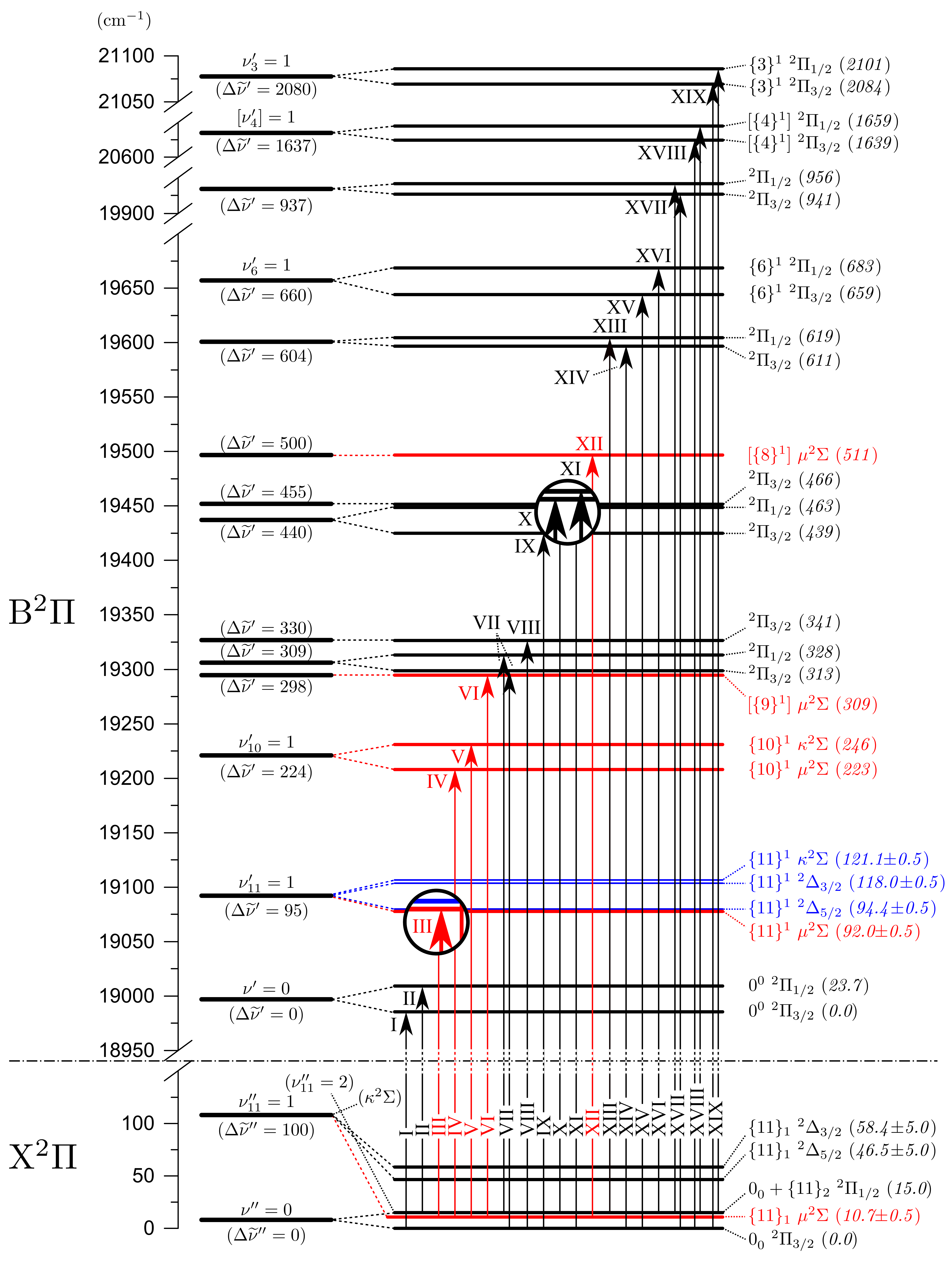}
  \caption{Energy level diagram for \ce{C6H}. The line cutting across the diagram separates the ground X$^2\Pi$ electronic state from the excited B$^2\Pi$ state. As plasma jets produce vibrationally cold molecules, all observed transitions originate from the lowest lying states: $0_0\ ^2\Pi_{3/2}$, $0_0\ ^2\Pi_{1/2}$, and $\{11\}_1\ \mu{^2}\Sigma$, with the $^2\Sigma - {^2}\Sigma$ transitions in red. Numbers in parentheses~() are the vibrational (left) and vibronic (right) frequencies expressed relative to the respective vibrational and vibronic ground states for each of the electronic states. States that are thus far unobserved experimentally are indicated in blue. Tentative mode identifications are given in square brackets~[\,]. Vibrational eigen modes are according to the numbering in Brown et al.~[\!\!\citenum{Brown1999}\,].}
  \label{fig:Energy-Level-Diagram}
\end{figure*}

The present study reports a comprehensive investigation of the vibronic structure of the B$^2\Pi -$X$^2\Pi$ transition in the \ce{C6H} molecule in the gas phase. Some 19 bands or band components have been detected, and in most cases rotationally resolved, for the main \ce{C6H} isotopologue, while eight have been recorded for \ce{^{13}C6H} and seven for \ce{C6D}. All assignments, taking into account all the arguments given here, result in a rather complete energy level diagram for \ce{C6H} (Fig.~\ref{fig:Energy-Level-Diagram}) in which the observed transitions (with corresponding band labeling) are shown. This figure includes energies pertaining to the \ce{C6H} main isotopologue; corresponding values for \ce{^{13}C6H} and \ce{C6D} can be obtained from Table~\ref{tab:summary}. For the non-degenerate ($\Sigma$) vibrations, besides the $0^0_0$ origin band three vibrational eigen modes have been identified: $\nu_6$, $\nu_3$, and, tentatively, $\nu_4$. In addition, a number of excited $^2\Pi$ vibronic states have been assigned to be probed in combination bands involving two low-lying degenerate vibrations. Low-lying bending modes have been observed for which a Renner-Teller treatment has been applied unambiguously, leading to the assignment of the $\mu^2\Sigma$ component of the $\nu_{11}$ mode and both $\mu^2\Sigma$ and $\kappa^2\Sigma$ components of the $\nu_{10}$ mode, while the $\mu^2\Sigma$ components of $\nu_9$ and $\nu_8$ have been tentatively assigned. These assignments are based on \emph{ab initio} calculations of the \ce{C6H} molecule, which were however carried out for the ground state electronic configuration of the molecule.~\cite{Liu1992,Brown1999,Cao2001} Previous matrix isolation data~\cite{Forney1995,Freivogel1995} and higher temperature discharge spectra~\cite{Kotterer1997} were useful in the analysis, as were the present results on the \ce{^{13}C6H} and \ce{C6D} isotopologues. The analysis and descriptive detail of the level structure of \ce{C6H} now equals that of the C$_4$H molecule~\cite{Hoshina1998} which is part of the same C$_n$H carbon chain series. 

The present comprehensive experimental study might be a stimulus for future \emph{ab initio} calculations specifically focusing on the B$^2\Pi$ electronically excited state of \ce{C6H}. Some open issues remain; those are the quest for a definite assignment of some vibrational (combination) modes, perturbation effects, and the dynamical effect of internal conversion in the highest vibrational levels. The investigation of the optical spectra of carbon chain molecules, and in particular also the \ce{C6H} molecule, was originally driven by the hypothesis that $^2\Pi - {^2}\Pi$ electronic transitions of C$_n$H radicals might be responsible for some of the diffuse interstellar bands~\cite{Fulara1993b} observed in the optical spectra of reddened stars. For the presently observed manifold of vibronic bands this is not the case.

\balance

\subsection*{Acknowledgements}
This work has been supported by the Netherlands Organization for Scientific Research (NWO) through a VICI grant and was performed within the context of the Dutch Astrochemistry Network (DAN).

\centering\rule[-5mm]{3cm}{1pt}



\bibliographystyle{achemso} 

\vspace{20pt}
\bibliography{Carbon-Chains-refs} 

\end{document}